\documentclass[preprint,aps,12pt,showpacs,nofootinbib,tightenlines]{revtex4}
\usepackage{mathrsfs}
\usepackage{amsmath}
\usepackage{amssymb}
\usepackage{CJK}
\usepackage{epsfig}
\usepackage{graphicx}
\usepackage{subfigure}
\usepackage{slashed}
\let\jnfont=\rm
\def\NPB#1,{{\jnfont { \it Nucl.\ Phys.\ B }}{\bf #1},}
\def\PLB#1,{{\jnfont { \it Phys.\ Lett.\ B }}{\bf #1},}
\def\EPJC#1,{{\jnfont {\it Eur.\ Phys.\ Jour.\ C }}{\bf #1},}
\def\EPL#1,{{\jnfont { \it Europhys.\ Lett.\ }}{\bf #1},}
\def\PR#1,{{\jnfont {\it Phys.\ Rept. }}{\bf #1},}
\def\PRD#1,{{\jnfont {\it Phys.\ Rev.\ D }}{\bf #1},}
\def\ZPC#1,{{\jnfont {\it Z.\  Phys.\ C }}{\bf #1},}
\def\PRL#1,{{\jnfont { \it Phys.\ Rev.\ Lett.\ }}{\bf #1},}
\def\MPLA#1,{{\jnfont{ \it  Mod.\ Phys.\ Lett\ A }}{\bf #1},}
\def\IJMPA#1,{{\jnfont{ \it  Int.\ J.\ Mod.\ Phys.\ A }}{\bf #1},}
\def\JPG#1,{{\jnfont {\it J.\ Phys.\ G\ }}{\bf #1},}
\def\CTP#1,{{\jnfont{ \it Commun.\ Theor.\ Phys.\ }}{\bf #1},}
\def\SCG#1,{{\jnfont{ \it Sci.\  China.\ G\ }}{\bf #1},}
\def\CPL#1,{{\jnfont{ \it Chin.\ Phys.\ Lett.\ }}{\bf #1},}
\def\CPC#1,{{\jnfont{ \it Chin.\ Phys.\ C\ }}{\bf #1},}
\def\JHEP#1,{{\jnfont {\it JHEP \ }}{\bf #1},}
\def\NPPS#1,{{\jnfont {\it Nucl.\ Phys.\ Proc.\ Suppl.\ }}{\bf #1},}

\def\pslash{\rlap{\hspace{0.02cm}/}{p}}
\begin{document}
\def\pslash{\rlap{\hspace{0.02cm}/}{p}}
\def\eslash{\rlap{\hspace{0.02cm}/}{e}}
\title {Higgs boson production in the $U(1)_{B-L}$ model at the ILC}
\author{Jinzhong Han$^{1}$}\email{hanjinzhong@zknu.edu.cn}
\author{Bingfang Yang$^{2,3}$}\email{yangbingfang@htu.edu.cn}
\author{Ning Liu$^{2}$}\email{wlln@mail.ustc.edu.cn}
\author{Jitao Li$^{1}$}
\affiliation{\footnotesize $^1$School of Physics and
Telecommunications Engineering, Zhoukou Normal University, Henan, 466001, China\\
 $^2$College of Physics and Electronic Engineering, Henan Normal University, Xinxiang 453007,
 China\\
 $^3$School of Materials Science and Engineering, Henan Polytechnic University,
Jiaozuo, 454000, China
   \vspace*{1.5cm}  }
\begin{abstract}

In the framework of the minimal $U(1)_{B-L}$ extension of the
Standard Model, we investigate the Higgs boson production processes
$e^{+}e^{-}\rightarrow ZH$, $e^{+}e^{-}\rightarrow
\nu_{e}\bar{\nu_{e}}H$, $e^{+}e^{-}\rightarrow t\bar{t}H$,
$e^{+}e^{-}\rightarrow ZHH$ and $e^{+}e^{-}\rightarrow
\nu_{e}\bar{\nu_{e}}HH$ at the International Linear Collider (ILC).
We present the production cross sections, the relative corrections
and compare our results with the expected experimental accuracies
for Higgs decay channel $H\rightarrow b\bar{b}$. In the allowed
parameter space, we find that the effects of the three single Higgs
boson production processes might approach the observable threshold
of the ILC. But the Higgs signal strengths $\mu_{b\bar{b}}$ of the
two double Higgs boson production processes are all out of the
observable threshold so that these effects will be difficult to be
observed at the ILC.

\end{abstract}
 \maketitle
\section{ Introduction}
\noindent

In the summer of 2012,  a bosonic resonance with a mass around 125
GeV was found at the Large Hadron Collider (LHC) by the ATLAS and
CMS Collaborations \cite{ATLAS-H,CMS-H}. So far, its properties are
compatible with the predictions of the Standard Model (SM) Higgs
boson. Meanwhile, the current LHC data is limited, there are still
large uncertainties about the couplings between the Higgs boson and
the other SM particles \cite{LHC-1,LHC-2,LHC-3,LHC-4,LHC-5}. Due to
the complicated background, the precision measurements of the
properties of the Higgs boson at the LHC are severely challenged. By
contrast, the Higgs factories beside the LHC, such as the
International Linear Collider (ILC) \cite{ILC-1,ILC-2,ILC-3}, can
measure the Higgs boson with high accuracy. In many cases, the ILC
can significantly improve the LHC measurements due to its clean
environment.

The ILC technical design report has pointed that it is planed to
measure Higgs boson at three center-of-mass (c.m.) energy: 250 GeV,
500 GeV and 1000 GeV.  In the first stage for $\sqrt{s}=250$ GeV,
the precision Higgs program will start at the Higgs-strahlung
process $e^{+}e^{-}\rightarrow ZH$, the cross section for this
process is dominant at the low energy and has the maximum cross
section  at around $\sqrt{s}=250$ GeV. In the second stage for
$\sqrt{s}=500$ GeV, the two very important processes
$e^{+}e^{-}\rightarrow t\bar{t}H$
and $e^{+}e^{-}\rightarrow ZHH$ are become accessible. For the
process $e^{+}e^{-}\rightarrow t\bar{t}H$, in which the top Yukawa
coupling appears at the tree-level for the first time at the ILC, it
will play an important role for the precision measurements of the
top quark Yukawa coupling. For the process $e^{+}e^{-}\rightarrow
ZHH$, to which the triple Higgs boson coupling contributes at the
tree-level, it will be crucial to understand the Higgs self-coupling
and the electroweak symmetry breaking. In the third stage for
$\sqrt{s}=1000$ GeV, the processes $e^{+}e^{-}\rightarrow
t\bar{t}H$, $e^{+}e^{-}\rightarrow \nu_{e}\bar{\nu_{e}}H$ and
$e^{+}e^{-}\rightarrow  \nu_{e}\bar{\nu_{e}}HH$ are involved. In
such energy stages, the channels $t\bar{t}H$ and
$\nu_{e}\bar{\nu_{e}}H$ have large cross section, and the channel
$\nu_{e}\bar{\nu_{e}}HH$ can be used together with the $ZHH$ process
to improve the measurement of the Higgs self-coupling. So far, many
relevant works mentioned above have been extensively studied in the
context of the SM
\cite{SM-zh-vvh-eeh-1,SM-zh-vvh-eeh-2,SM-zh-vvh-eeh-3,SM-zh-vvh-eeh-4,SM-zh-vvh-eeh-5,SM-zh-vvh-eeh-6,SM-zh-vvh-eeh-7}
and some new physics models
\cite{np-1,np-2,np-3,np-4,np-5,np-6,np-7,np-8,np-9,np-10,np-11,np-12,np-13,np-14,np-15,np-16,np-17,np-18,np-19,np-20}.

The minimal $B-L$ extension of the SM is based on the structure
$SU(3)_C\times SU(2)_L\times U(1)_Y\times U(1)_{B-L}$ gauge
symmetry, in which the SM gauge has a further $U(1)_{B-L}$ group
related to the Baryon minus Lepton $(B-L)$ gauged number
\cite{B-L-1,B-L-2}. It was known that this model is in agreement
with the current experimental results of the light neutrino masses
and their large mixing. The $B-L$ model predicted some new particles
beyond the SM, such as the new heavy gauge bosons, the heavy
neutrino and the heavy neutral Higgs boson. In addition, some
couplings of the Higgs boson in the $B-L$ model are modified with
respect to the SM. These new effects will alter the property of the
SM Higgs boson and influence various SM Higgs boson processes,
making the model phenomenologically rich and testable at the LHC and
the ILC
\cite{B-L-2,B-L-LHC-ILC-1,B-L-LHC-ILC-2,B-L-LHC-ILC-3,B-L-LHC-ILC-4,
B-L-LHC-ILC-5,B-L-LHC-ILC-6,B-L-LHC-ILC-7,B-L-LHC-ILC-8,B-L-LHC-ILC-9,B-L-LHC-ILC-10}.
In this paper, we mainly study the single Higgs boson production
processes $e^{+}e^{-}\rightarrow ZH$, $e^{+}e^{-}\rightarrow
\nu_{e}\bar{\nu_{e}}H$, $e^{+}e^{-}\rightarrow e^{+}e^{-}H$,
$e^{+}e^{-}\rightarrow t\bar{t}H$ and the double Higgs boson
production processes $e^{+}e^{-}\rightarrow ZHH$,
$e^{+}e^{-}\rightarrow \nu_{e}\bar{\nu_{e}}HH$ in the $B-L$ model at
the ILC.

The paper is organized as follows. In Sec.II we briefly review the
basic content of the $B-L$ model related to our work. In Sec.III and
Sec.IV we respectively investigate the Higgs boson production
processes and the Higgs signal strengths in the $B-L$ model at the
ILC. Finally, we give a summary in Sec.V.

\section{ A brief review of the B-L model}
Here we will briefly review the ingredients relevant to our
calculations, the detailed description of the $B-L$ model can be
found in Refs. \cite{B-L-LHC-ILC-1,B-L-LHC-ILC-4}. The $B-L$ model
is the minimal extensions of the SM
\cite{B-L-extension-1,B-L-extension-2,B-L-extension-3,B-L-extension-4,B-L-extension-5}
with the classical conformal symmetry, and based on the gauge group
$SU(3)_C\times SU(2)_L\times U(1)_Y\times U(1)_{B-L}$. The
Lagrangian for the fermionic and kinetic sectors  are given by
\begin{eqnarray}
\mathcal{L}_{B-L} &=&i~\bar{l}D_{\mu }\gamma ^{\mu
}l+i~\bar{e}_{R}D_{\mu }\gamma ^{\mu }e_{R}+i~\bar{\nu}_{R}D_{\mu
}\gamma ^{\mu }\nu _{R}  \nonumber
\\
&&-\frac{1}{4}W_{\mu \nu }W^{\mu \nu }-\frac{1}{4}B_{\mu \nu }B^{\mu \nu }-%
\frac{1}{4}C_{\mu \nu }C^{\mu \nu }.
\end{eqnarray}%
The covariant derivative $D_{\mu }$ is different from the SM one by
the term $ig^{\prime }Y_{B-L}C_{\mu }$, where $g^{\prime}$ is the $%
U(1)_{B-L}$ gauge coupling constant, $Y_{B-L}$ is the $B-L$ charge, and $%
C_{\mu \nu }=\partial _{\mu }C_{\nu }-\partial _{\nu }C_{\mu }$ is
the field strength of the $U(1)_{B-L}$.

The Lagrangian for the Higgs and Yukawa sectors are given by
\begin{eqnarray}
\mathcal{L}_{B-L} &=&(D^{\mu }\phi )(D_{\mu }\phi )+(D^{\mu }\chi
)(D_{\mu
}\chi )-V(\phi ,\chi )  \nonumber \\
&&-\Big(\lambda _{e}\bar{l}\phi e_{R}+\lambda _{\nu
}\bar{l}\tilde{\phi}{\nu
}_{R}+\frac{1}{2}\lambda _{\nu _{R}}\bar{\nu ^{c}}_{R}\chi \nu _{R} +h.c.%
\Big).
\end{eqnarray}%
The $U(1)_{B-L}$ and $SU(2)_{L}\times U(1)_{Y}$ gauge symmetries can
be spontaneously broken by a SM singlet complex scalar field $\chi $
and a complex $SU(2)$ doublet of scalar fields $\phi $,
respectively.

The scalar potential $V(\phi ,\chi )$ is
given by%
\begin{eqnarray}
V(\phi ,\chi ) &=&m_{1}^{2}\phi ^{\dagger }\phi +m_{2}^{2}\chi
^{\dagger }\chi +\lambda _{1}(\phi ^{\dagger }\phi )^{2}+\lambda
_{2}(\chi ^{\dagger }\chi )^{2}+\lambda _{3}(\chi ^{\dagger }\chi
)(\phi ^{\dagger }\phi ). \label{scpot}
\end{eqnarray}
To determine the condition for the potential to be bounded from
below, the couplings $\lambda_{1},\lambda_{2}$ and $\lambda_{3}$
should be related with
$ 4\lambda _{1}\lambda _{2}-\lambda _{3}>0,
\lambda _{1}\geq 0,\lambda _{2}\geq 0$. The vev's, $|\langle \phi
\rangle |=v/\sqrt{2}$ and $|\langle \chi \rangle |=v^{\prime
}/\sqrt{2}$, are then given by
\begin{eqnarray}
v^{2}=\frac{4\lambda _{2}m_{1}^{2}-2\lambda _{3}m_{2}^{2}}{\lambda
_{3}^{2}-4\lambda _{1}\lambda _{2}},~\ \ \ \ ~~v^{\prime 2}=\frac{%
-2(m_{1}^{2}+\lambda _{1}v^{2})}{\lambda _{3}}.
\end{eqnarray}%
where $v$ and $v^{\prime}$ are the electroweak symmetry breaking
scale and the $B-L$ symmetry breaking scale, respectively.

After the electroweak symmetry breaking, one obtains the mass  of
the gauge bosons
\begin{align}
m_{\gamma} &=0, \nonumber \\
 m_{W^{\pm}} &= \frac{1}{2} v g, \nonumber \\
 m_Z &= \frac{v}{2}\sqrt{g^2 + g_1^2}, \nonumber \\
 m_{Z'} &=  2 v' g'.
\end{align}
where $g$ and $g_1$ are the $SU(2)_L$ and $U(1)_Y$ gauge couplings.
The $Z^{\prime}$ boson mass is constrained from the most recent
limit at LEP \cite{LEP-constrain}
\begin{equation} m_{Z^{\prime }}/g^{\prime}>7~ {\rm TeV}.
\end{equation}%

The mixing between the SM complex $SU(2)_L$ doublet and complex
scalar singlet is controlled by the coupling $\lambda_3$ as shown in
Eq. (\ref{scpot}). This mixing can be expressed by the mass matrix
$\phi$ and $\chi$
\begin{equation}
\frac{1}{2}m^{2}(\phi ,\chi )=\left(
\begin{array}{cc}
\lambda _{1}v^{2} & \frac{\lambda _{3}}{2}vv^{\prime } \\
\frac{\lambda _{3}}{2}vv^{\prime } & \lambda _{2}v^{\prime 2}%
\end{array}%
\right) .
\end{equation}%
Therefore, the mass eigenstates fields $H$ and $H^{\prime }$ are
given by
\begin{equation}
\left(
\begin{array}{c}
H \\
H^{\prime }%
\end{array}%
\right) =\left(
\begin{array}{cc}
\cos \alpha  & -\sin \alpha  \\
\sin \alpha  & \cos \alpha
\end{array}%
\right) \left(
\begin{array}{c}
\phi  \\
\chi
\end{array}%
\right) ,
\end{equation}%
where the mixing angle $\alpha $ is defined by
\begin{equation}\label{tan2a}
\tan 2\alpha =\frac{|\lambda _{3}|vv^{\prime }}{\lambda
_{1}v^{2}-\lambda _{2}v^{\prime 2}}.
\end{equation}%
The masses of $H$ and $H^{\prime }$ are given by
\begin{equation}\label{mhh}
m_{H, H^{\prime }}^{2}=\lambda _{1}v^{2}+\lambda _{2}v^{\prime 2}\mp \sqrt{%
(\lambda _{1}v^{2}-\lambda _{2}v^{\prime 2})^{2}+\lambda
_{3}^{2}v^{2}v^{\prime 2}}.
\end{equation}%
Here, $H$ and $H^{\prime }$ are light and heavy Higgs bosons,
respectively.

From Eqs.~(\ref{tan2a}) and (\ref{mhh}), it is straightforward to
have:
\begin{eqnarray}\label{isomorphism}
\lambda_1&=& \frac{ m_{H}^2}{2v^2}\cos^2{\alpha} + \frac{m_{H'}^2}{2
v^2}\sin^2{\alpha},\nonumber \\
\lambda_2&=& \frac{ m_{H}^2}{2{v'}^2}\sin^2{\alpha} +
\frac{m_{H'}^2}{2
{v'}^2}\cos^2{\alpha}, \nonumber \\
\lambda_3&=&\frac{ \left( m_{H'}^2 - m_{H}^2 \right)}{
2vv'}\sin{(2\alpha)}.
\end{eqnarray}

Because of the mixing between the two Higgs bosons $H$ and $H'$, the
usual couplings among the SM-like Higgs $H $ boson and the SM
particles are modified. Additionally, there are new couplings among
the extra Higgs $H^{\prime} $ and the SM particles, which will lead
to a different Higgs phenomenology from the SM. Notice that the
scalar mixing angle $\alpha $ is a free parameter of the model, and
the light(heavy) Higgs boson couples to the new matter content
proportionally to sin$\alpha$ (cos$\alpha$). The relevant Feynman
rules involved in our calculations are given in Table A.1 of App. A,
which can be found in Refs. \cite{B-L-2, B-L-LHC-ILC-7}.

\section{Higgs productions in the B-L model at ILC }
In our numerical calculations,  we take the SM  parameters as:
$m_t=$172.4 GeV, ${\sin}^{2}\theta_{W}=$0.23126, $m_{Z}=$91.187 GeV,
$m_{H}=$125 GeV, $\alpha(m_Z)$=1/128
 \cite{PDG-2014}. For
the $B-L$ parameters, the mixing angle $\alpha$, the gauge coupling
constant $g'$, the mixing gauge coupling $\tilde{g}$, the masses
$m_{Z'}$, $m_{H'}$ and $m_{\nu_H}$ are involved. The
Ref.\cite{B-L-consition-1} has discussed the constraints on these
parameters from experiment and theory, and points out that ${\rm
sin} \alpha\leq 0.36$,  $m_{Z'}\geq1830$ GeV, $m_{\nu_H}\sim500~{\rm
GeV}$, $m_{H'}\geq 125~{\rm GeV}$. In the following calculations, we
vary ${\rm sin} \alpha$ in the range of $0.05 \leq {\rm
sin}\alpha\leq 0.4$, and take $m_{Z'}=  2500~{\rm GeV}$,
$m_{\nu_H}=500~{\rm GeV}$, $m_{H'}=500~{\rm GeV}$, $g'=0.3$,
$\tilde{g}=-0.1$. All the numerical results are done by using
\textsf{CalcHEP 3.6.25} package \cite{calchep}.
\subsection{Single Higgs boson productions}
\noindent

\begin{figure}[htbp]
\scalebox{0.53}{\epsfig{file=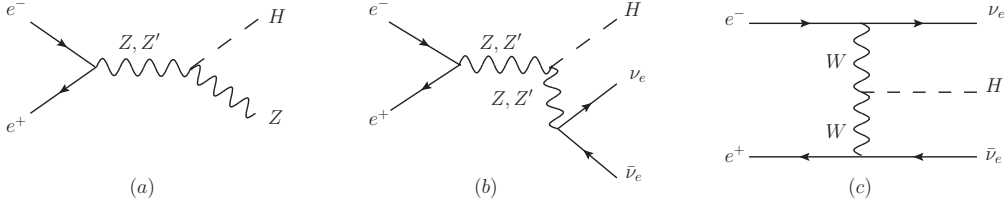}}\hspace{0cm} \vspace{-0.5cm}
\caption{Lowest-order Feynman diagrams for $e^{+}e^{-}\rightarrow
ZH$(a) and $e^{+}e^{-}\rightarrow \nu_{e}\bar{\nu_{e}}H$(b,c) in the
$B-L$ model.}\label{fig:eezhvvhfn}
\end{figure}

\begin{figure}[htbp]
\scalebox{0.53}{\epsfig{file=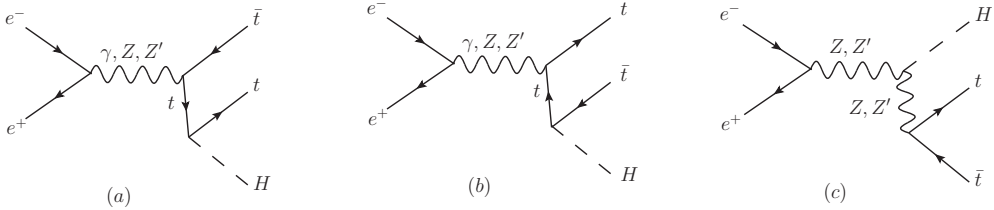}}\vspace{-0.5cm}
\caption{Lowest-order Feynman diagrams for $e^{+}e^{-}\rightarrow
t\bar{t}H$ in the $B-L$ model.} \label{fig:eetthfn}
\end{figure}

In Fig.\ref{fig:eezhvvhfn} and Fig.\ref{fig:eetthfn}, we show the
lowest-order Feynman diagrams of the single Higgs boson production
processes $e^{+}e^{-}\rightarrow ZH$, $e^{+}e^{-}\rightarrow
\nu_{e}\bar{\nu_{e}}H$ and $e^{+}e^{-}\rightarrow t\bar{t}H$ in the
$B-L$ model. In comparison with the SM, we can see that these three
processes receive the additional contributions from the heavy gauge
boson $Z'$ and the modified couplings of $HXX$ at the tree-level.

\begin{figure}[htbp]
\begin{center}
\scalebox{0.3}{\epsfig{file=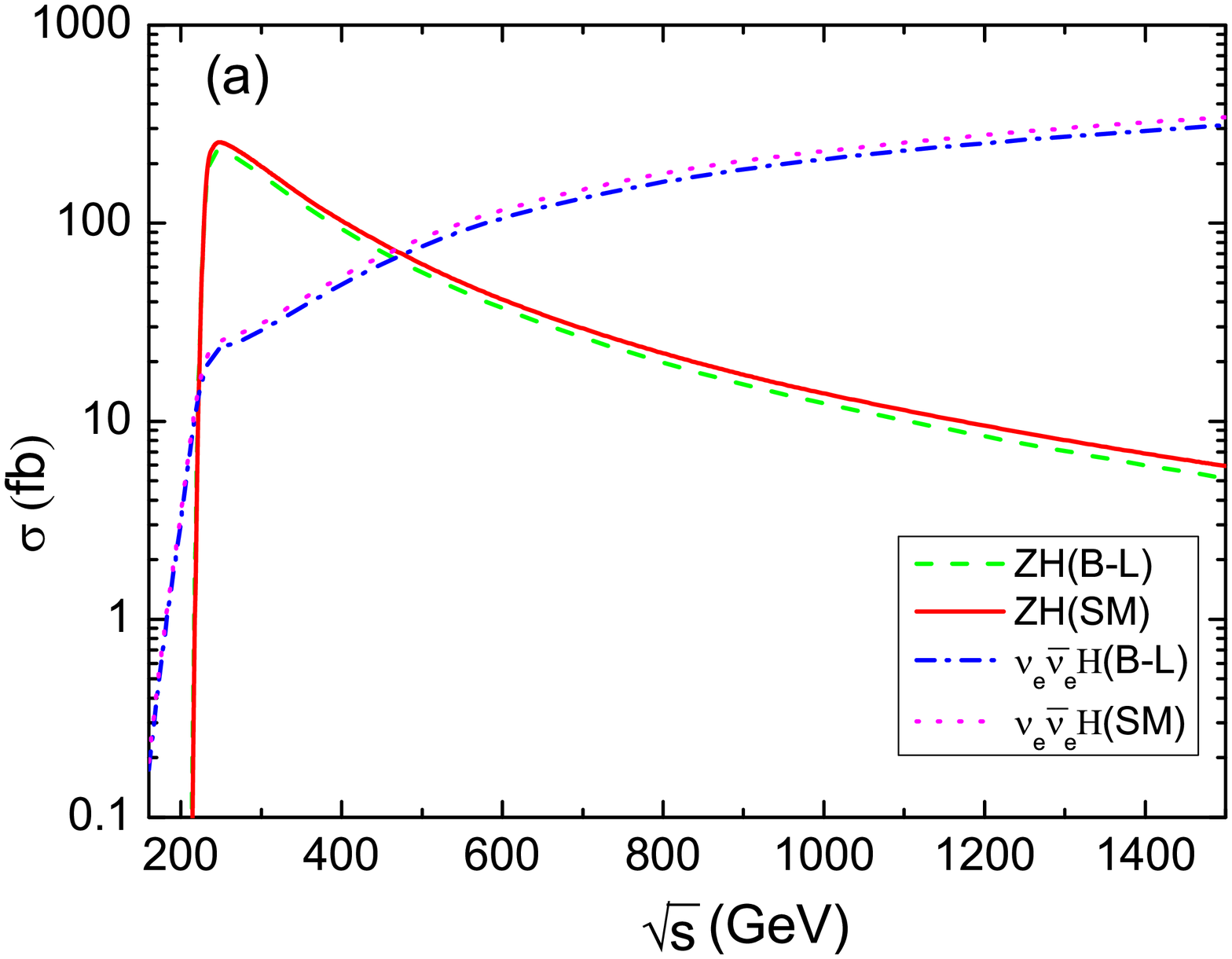}}\hspace{-1.5cm}
\scalebox{0.3}{\epsfig{file=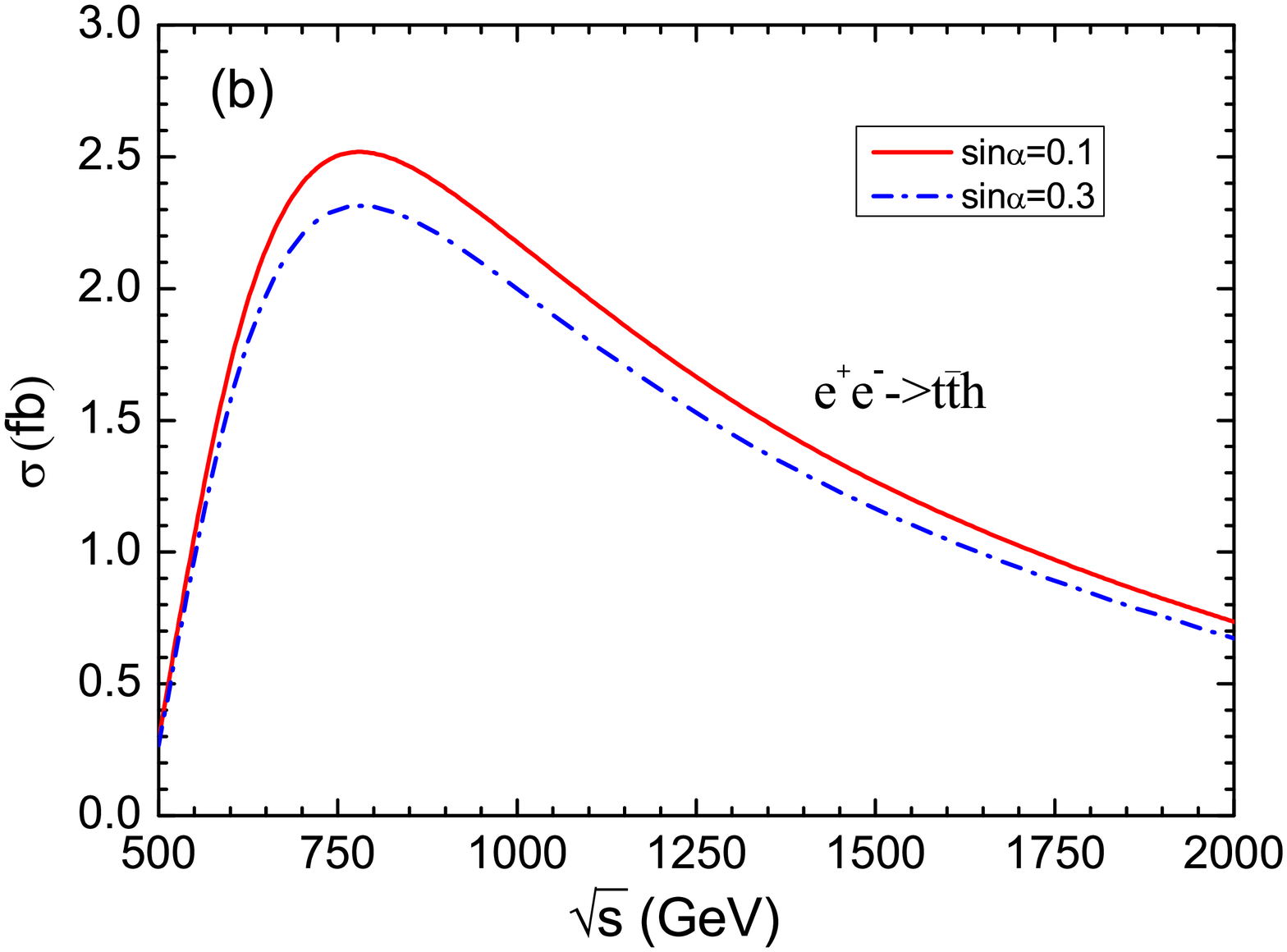}}\hspace{-1cm}
\vspace{-0.5cm}
 \caption{The
production cross section $\sigma$ for the process
$e^{+}e^{-}\rightarrow ZH$, $e^{+}e^{-}\rightarrow
\nu_{e}\bar{\nu_{e}}H$ (a) and $e^{+}e^{-}\rightarrow t\bar{t}H$ (b)
versus the c.m. energy $\sqrt{s}$ in the SM and $B-L$ model.
}
\label{fig:zhvvhtths}
\end{center}
\end{figure}

In Fig.\ref{fig:zhvvhtths}, we show the production cross sections
$\sigma$ of these three processes versus the c.m. energy $\sqrt{s}$
in the SM and $B-L$ model, respectively. We can see that the process
$e^{+}e^{-}\rightarrow ZH$ reaches its maximum at $\sim$ 250 GeV.
The $\nu_{e}\bar{\nu_{e}}H$ production cross sections increase with
the $\sqrt{s}$ and can take over that of the $ZH$ process at
$\sqrt{s}\geq 500$ GeV. Similar to the process
$e^{+}e^{-}\rightarrow ZH$, the $t\bar{t}H$ production cross
sections increase firstly and then decrease with the $\sqrt{s}$ and
reaches its maximum at $\sim$ 800 GeV.  The cross sections of these
three production processes in the $B-L$ model are all lower than
their SM values.

Considering the polarization of the initial electron and positron
beams, the cross section of a process can be expressed as
\cite{polarization-1,polarization-2}
\begin{eqnarray}
\sigma(P_{e^{-}},P_{e^{+}})&=& \frac{1}{4}[(1+P_{e^{-}})(1+P_{e^{+}})\sigma_{RR}+(1-P_{e^{-}})(1-P_{e^{+}})\sigma_{LL} \nonumber \\
 &&
 +(1+P_{e^{-}})(1-P_{e^{+}})\sigma_{RL}+(1-P_{e^{-}})(1+P_{e^{+}})\sigma_{LR}],
\end{eqnarray}
 where
$P_{e^{-}}$ and $P_{e^{+}}$ are the polarization degree of the
electron and positron beam, respectively. As in Ref. \cite{ILC-2},
we assume $P(e^{-},e^{+})=(-0.8,0.3)$ at $\sqrt{s}$=250, 500 GeV and
$P(e^{-},e^{+})=(-0.8,0.2)$ at $\sqrt{s}$=1000 GeV  in our
calculations.

\begin{figure}[htbp]
\scalebox{0.3}{\epsfig{file=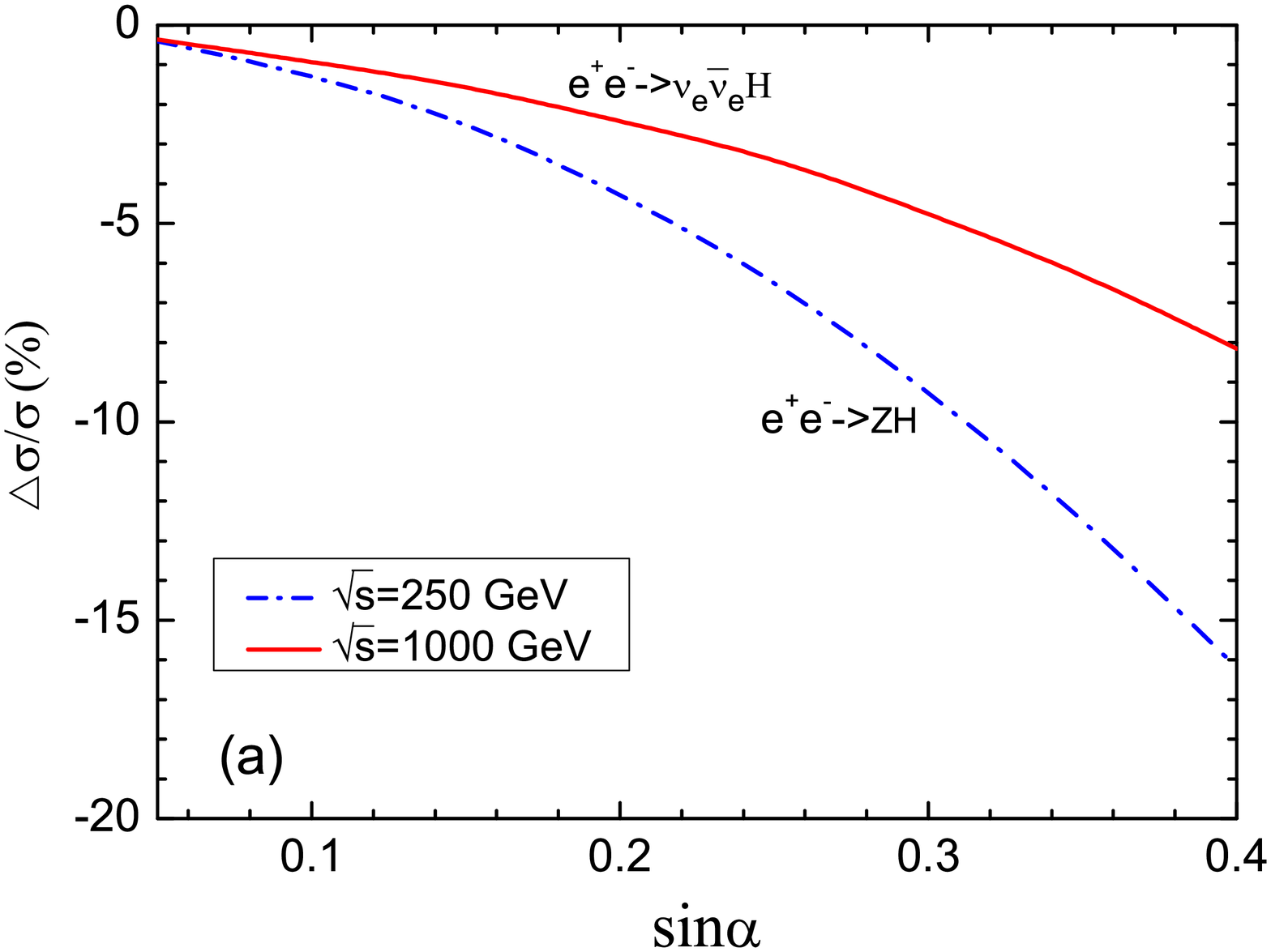}}\hspace{-1.5cm}
\scalebox{0.3}{\epsfig{file=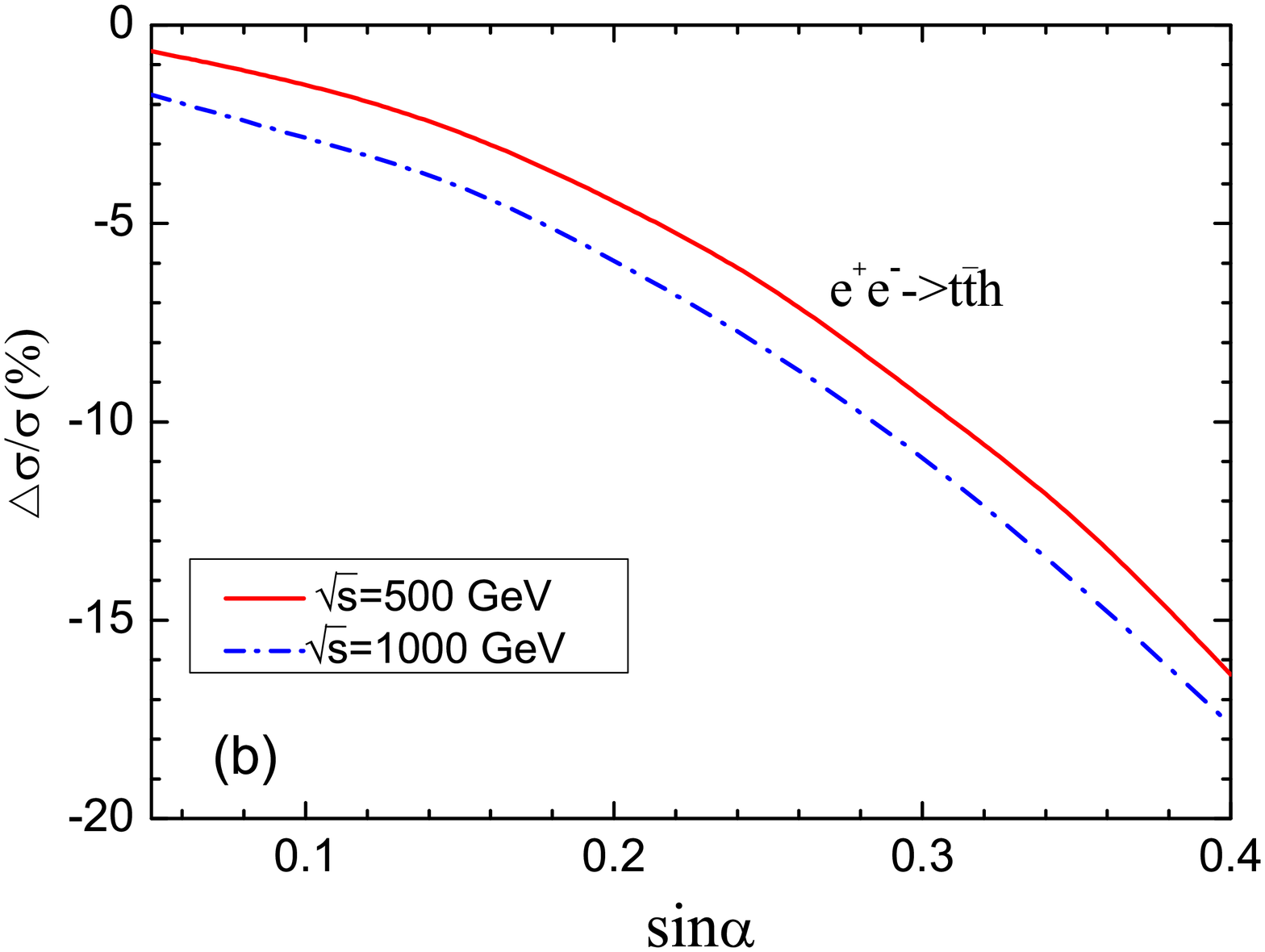}}\vspace{-0.7cm} \caption{The
relative correction $\Delta\sigma/\sigma$ for the process
$e^{+}e^{-}\rightarrow ZH$, $e^{+}e^{-}\rightarrow
\nu_{e}\bar{\nu_{e}}H$ (a) and $e^{+}e^{-}\rightarrow t\bar{t}H$ (b)
versus  ${\rm sin}\alpha$  for different c.m. energy $\sqrt{s}$ in
the $B-L$ model.}\label{fig:delta-zh-vvh}
\end{figure}

In Fig.\ref{fig:delta-zh-vvh}, we show the relative corrections
$\Delta\sigma/\sigma$=$(\sigma_{B-L}-\sigma_{SM})/\sigma_{SM} $ of
the three single Higgs boson production channels versus the mixing
angle ${\rm sin}\alpha$ for $\sqrt{s}=250, 500, 1000$ GeV at the ILC
with polarized beams. For these three processes, we can see that the
values of the relative corrections are all negative and increase
with the ${\rm sin}\alpha$ increasing, the $\Delta\sigma/\sigma$ of
processes $ZH$, $\nu_{e}\bar{\nu_{e}}H$, $t\bar{t}H$ can
respectively reach $-16.2\%, -8.2\%, -16.4\%$. Due to the fact that
the effects of the heavy gauge boson $Z'$ decouple, the relative
corrections $\Delta\sigma/\sigma$ are insensitive to the $m_{Z'}$,
so we do not show the dependence of the relative corrections on
$m_{Z'}$ here.

At the ILC with $\sqrt{s}$ = 250 GeV,  the total SM electroweak
correction for the $ZH$ production process is about 5\%
\cite{sm1,sm2}. Meanwhile, the ILC can measure the cross section for
$ZH$ and $\nu_{e}\bar{\nu_{e}}H$ to a relative accuracy of
$2.0-2.6\%$ and $2.2-11\%$ \cite{ILC-2}. At the ILC with $\sqrt{s}$
= 1000 GeV, the expected accuracies for $t\bar{t}H$ process may
achieve an even more remarkable precision of 6.3\%\cite{ILC-2}.
Thus,  the $B-L$ model effects on these three processes might be
observed at the ILC for the large sin$\alpha$.
\subsection{Double Higgs boson productions}

\begin{figure}[htbp]
\scalebox{0.41}{\epsfig{file=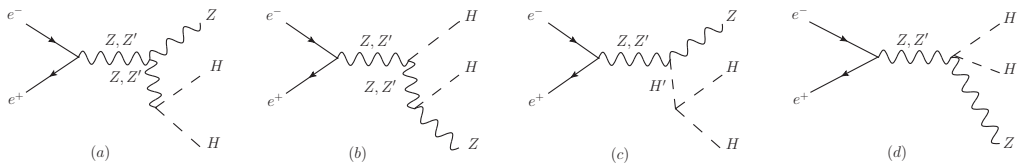}}\vspace{-0.5cm}
\caption{Lowest-order Feynman diagrams for $e^{+}e^{-}\rightarrow
ZHH$ in the $B-L$ model.}\label{fig:eezhhfn}
\end{figure}

\begin{figure}[htbp]
\scalebox{0.41}{\epsfig{file=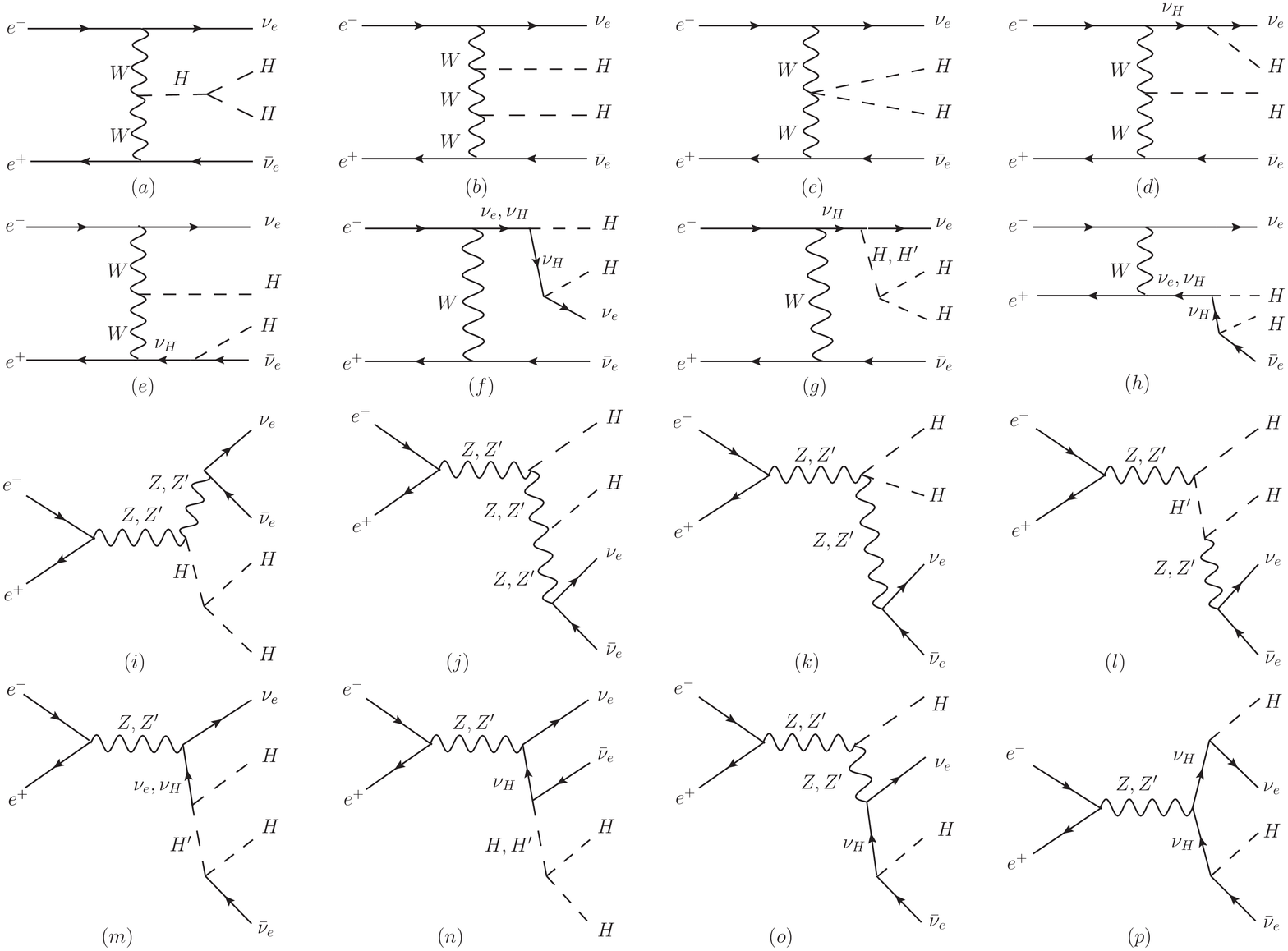}}\vspace{-0.5cm}
\caption{Lowest-order Feynman diagrams for $e^{+}e^{-}\rightarrow
\nu_{e}\bar{\nu_{e}}HH$ in the $B-L$ model.}\label{fig:eevvhhfn}
\end{figure}

At the ILC, the main triple Higgs boson coupling can be studied
through the double Higgs-strahlung off $Z$ boson process
$e^{+}e^{-}\rightarrow ZHH$ and double Higgs fusion process
$e^{+}e^{-}\rightarrow \nu_{e}\bar{\nu_{e}}HH$. The relevant Feynman
diagrams are shown in Fig.\ref{fig:eezhhfn} and
Fig.\ref{fig:eevvhhfn}.

In Fig.7(a), we show the cross sections for the two processes versus
the c.m. energy $\sqrt{s}$ in the SM and the $B-L$ model for
sin$\alpha$ = 0.3. We can see that the cross section for the process
$e^{+}e^{-}\rightarrow ZHH$ reaches its maximum at around 500 GeV.
It is noteworthy that the process $e^{+}e^{-}\rightarrow
\nu_{e}\bar{\nu_{e}}HH$ will become sizable at $\sqrt{s}$ = 1000 GeV
and can be used together with the $e^{+}e^{-}\rightarrow ZHH$
process to improve the measurement of the Higgs self-coupling.
Furthermore, we can see that the two processes have a similar trend
in the SM and the $B-L$ model.

\begin{figure}[htbp]
\scalebox{0.3}{\epsfig{file=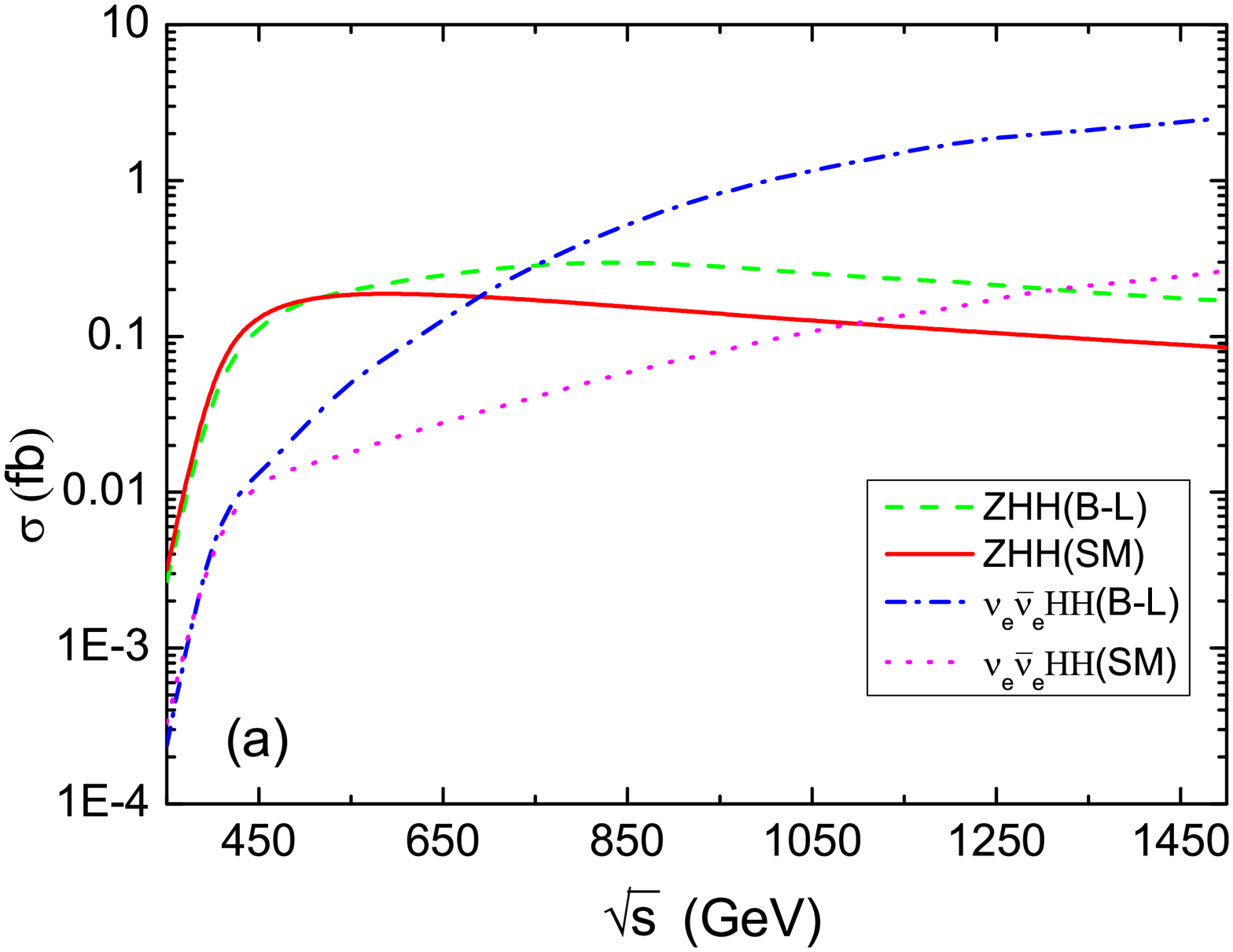}}\hspace{-1.5cm}
\scalebox{0.3}{\epsfig{file=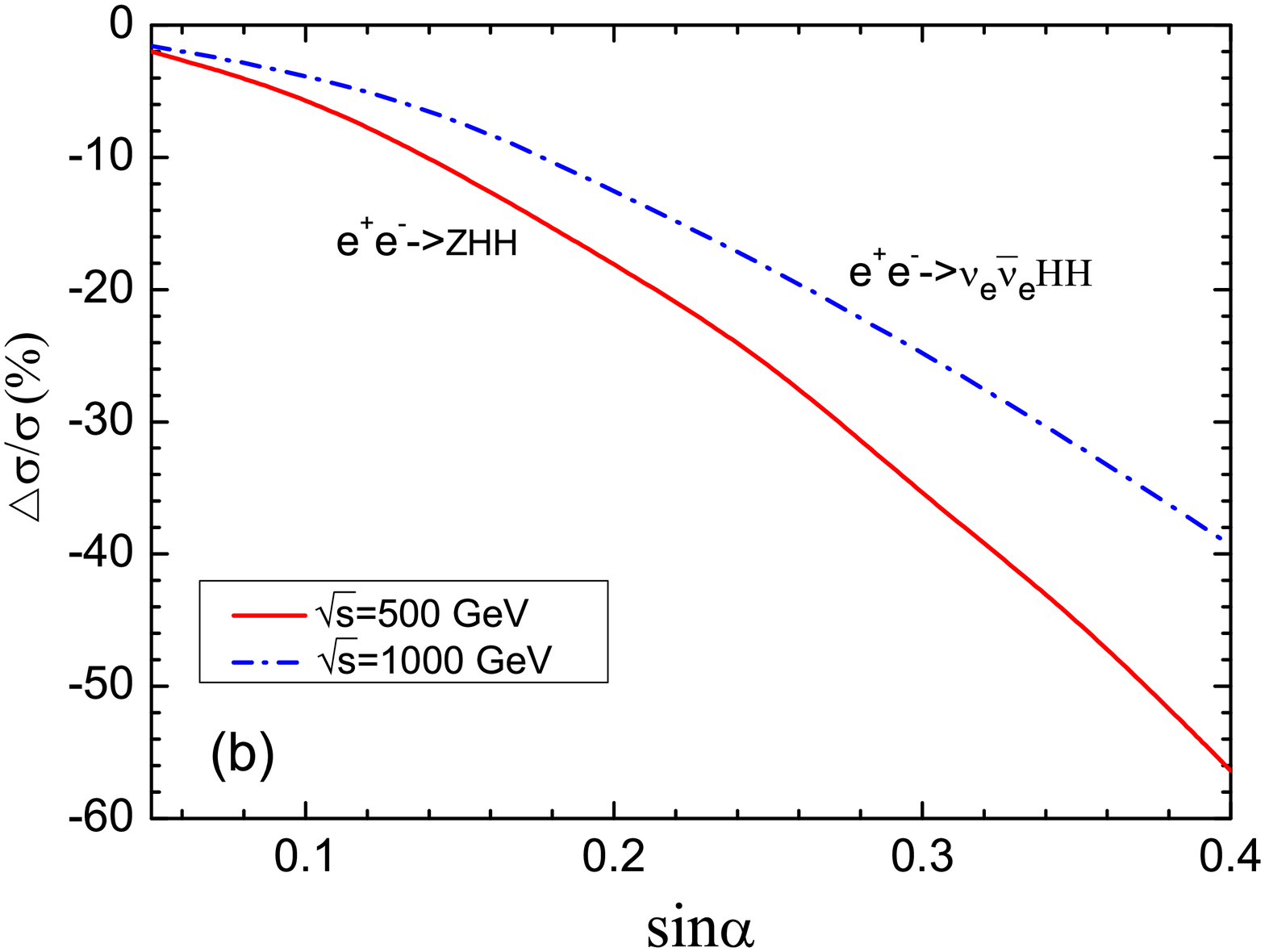}}\vspace{-0.7cm}
\caption{The production cross section $\sigma$ and the relative
correction $\Delta\sigma/\sigma$  for the processes
$e^{+}e^{-}\rightarrow ZHH$  and $e^{+}e^{-}\rightarrow
\nu_{e}\bar{\nu_{e}}HH$ versus the c.m. energy $\sqrt{s}$ (a) and
sin$\alpha$ (b) in the $B-L$ model.}\label{fig:zhhvvhhs}
\end{figure}

In Fig.7(b), we show the relative corrections of these two double
Higgs production processes versus ${\rm sin}\alpha$ for
$\sqrt{s}=500, 1000$ GeV with polarized beams at the ILC. We can see
that the relative corrections are negative and the values become
larger with the increasing of the ${\rm sin}\alpha$, which is
similar to the behavior of the single Higgs production processes
mentioned above. In the region of large sin$\alpha$, the
$\Delta\sigma/\sigma$ of processes $ZHH$ and
$\nu_{e}\bar{\nu_{e}}HH$ can reach $-56.4\%$ for $\sqrt{s}=500$ GeV
and $-39.3\%$ for $\sqrt{s}=1000$ GeV, respectively. The Refs.
\cite{LHC-1,LHC-2,LHC-5,HL-LHC-1,HL-LHC-2,HL-LHC-3,HL-LHC-4,HL-LHC-5,HL-LHC-6}
suggest that the expected accuracy for the $HHH$ coupling could be
reached 50\% through $pp\rightarrow HH\rightarrow bb\gamma\gamma$ at
the HL-LHC with $\mathcal {L}$=3000 fb$^{-1}$, and this accuracy may
be further improved to be around 13\% at the ILC with
$\sqrt{s}$=1000 GeV \cite{LHC-1,LHC-2,LHC-5,HL-LHC-1}. By this
token, the effects of the $B-L$ model might be observed through
these two processes at the ILC.

\section{The Higgs signal strengths in the B-L model}
In order to provide more information for probing the Higgs boson
processes, we give the Higgs signal strengths in the $B-L$ model.
Considering the Higgs boson decay mode, the signal strengths can be
defined as
\begin{eqnarray}
 \mu_{i}=\frac{\sigma_{ B-L}\times BR(H\rightarrow
i)_{ B-L}}{\sigma_{SM}\times BR(H\rightarrow i)_{SM}},
\end{eqnarray}
where $i$ denotes a possible final state of the SM fermion and boson
pairs.
\begin{table}[htbp]
\begin{center}
\caption{ Expected accuracies for cross section times branching
ratio measurements for the 125 GeV Higgs boson \cite{ILC-2}. }
\label{table1} \vspace{0.2cm}
\begin{tabular}{|c|c|c|c|c|c|c|c|c|c|}
 \hline  &\multicolumn{9}{c|}{ $\Delta(\sigma\cdot BR)/(\sigma\cdot BR)$}\\
 \hline $\mathcal {L}$ and $\sqrt{s}$&\multicolumn{2}{c|}{250 fb$^{-1}$ at 250 GeV}  &\multicolumn{4}{c|}{500 fb$^{-1}$ at 500 GeV }&\multicolumn{3}{c|}{1000 fb$^{-1}$ at 1000 GeV }\\
$(P_{e^{-}}, P_{e^{+}})$ &\multicolumn{2}{c|}{  (-0.8, +0.3) }&\multicolumn{4}{c|}{ (-0.8, +0.3)}&\multicolumn{3}{c|}{ (-0.8, +0.2)}\\
 \hline mode &$ZH$&$\nu_{e}\bar{\nu_{e}}H$& $ZH$ & $\nu_{e}\bar{\nu_{e}}H$ & $t\bar{t}H$ & $ZHH$ & $\nu_{e}\bar{\nu_{e}}H$ & $t\bar{t}H$ & $\nu_{e}\bar{\nu_{e}}HH$  \\
 \hline $H\rightarrow b\bar{b}$ & $1.1\%$ &$10.5\%$ & $1.8\%$ & $0.66\%$ & $35\%$& $64\%$ & $0.47\%$ & $8.7\%$ & $38\%$  \\
 \hline $H\rightarrow c\bar{c}$ & $7.4\%$ &-& $12\%$ & $6.2\%$ &-&-&$7.6\%$ &- &-\\
 \hline $H\rightarrow gg$ & $9.1\%$ &-& $14\%$ & $4.1\%$& - &-& $3.1\%$ &- &-\\
 \hline $H\rightarrow WW^{\ast}$ & $9.1\%$ &-& $9.2\%$ & $2.6\%$ &-& -& $3.3\%$ &- &-\\
 \hline $H\rightarrow \tau^{+}\tau^{-}$ & $4.2\%$ &-& $5.4\%$ & $14\%$ &-&-& $3.5\%$ &- &-\\
 \hline $H\rightarrow ZZ^{\ast}$ & $19\%$ &-& $25\%$ & $8.2\%$ &-& -&$4.4\%$ &- &-\\
 \hline $H\rightarrow \gamma\gamma$ & 29-38\% &-& 29-38\% & 20-26\%&-&-& 7-10\% &- &-\\ \hline
 \end{tabular}\end {center} \end{table}

The expected accuracies for $\Delta(\sigma\cdot BR)/(\sigma\cdot
BR)$ measurements for $m_{H}=125$ GeV at the ILC are shown in Table
\ref{table1}. Due to the $b\bar{b}$ decay mode is more easily
achievable than other modes \cite{ILC-1,ILC-2}, we only consider
this decay mode in the following section. The expected precision
limits of the $b\bar{b}$ mode respectively correspond to the blue
dash-dot lines in the numerical figures.
\begin{figure}[htbp]
\scalebox{0.3}{\epsfig{file=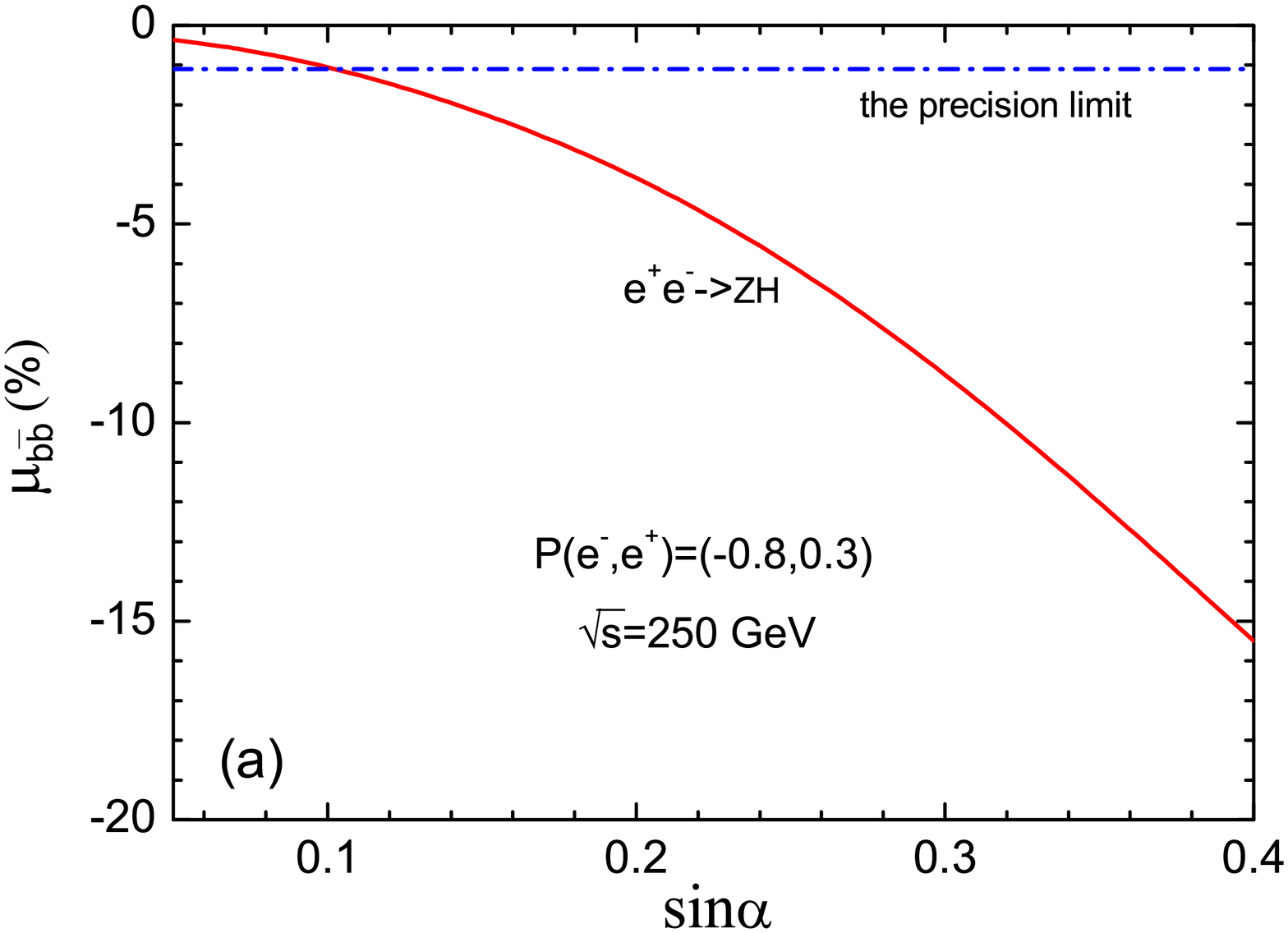}}\hspace{-1.5cm}
\scalebox{0.3}{\epsfig{file=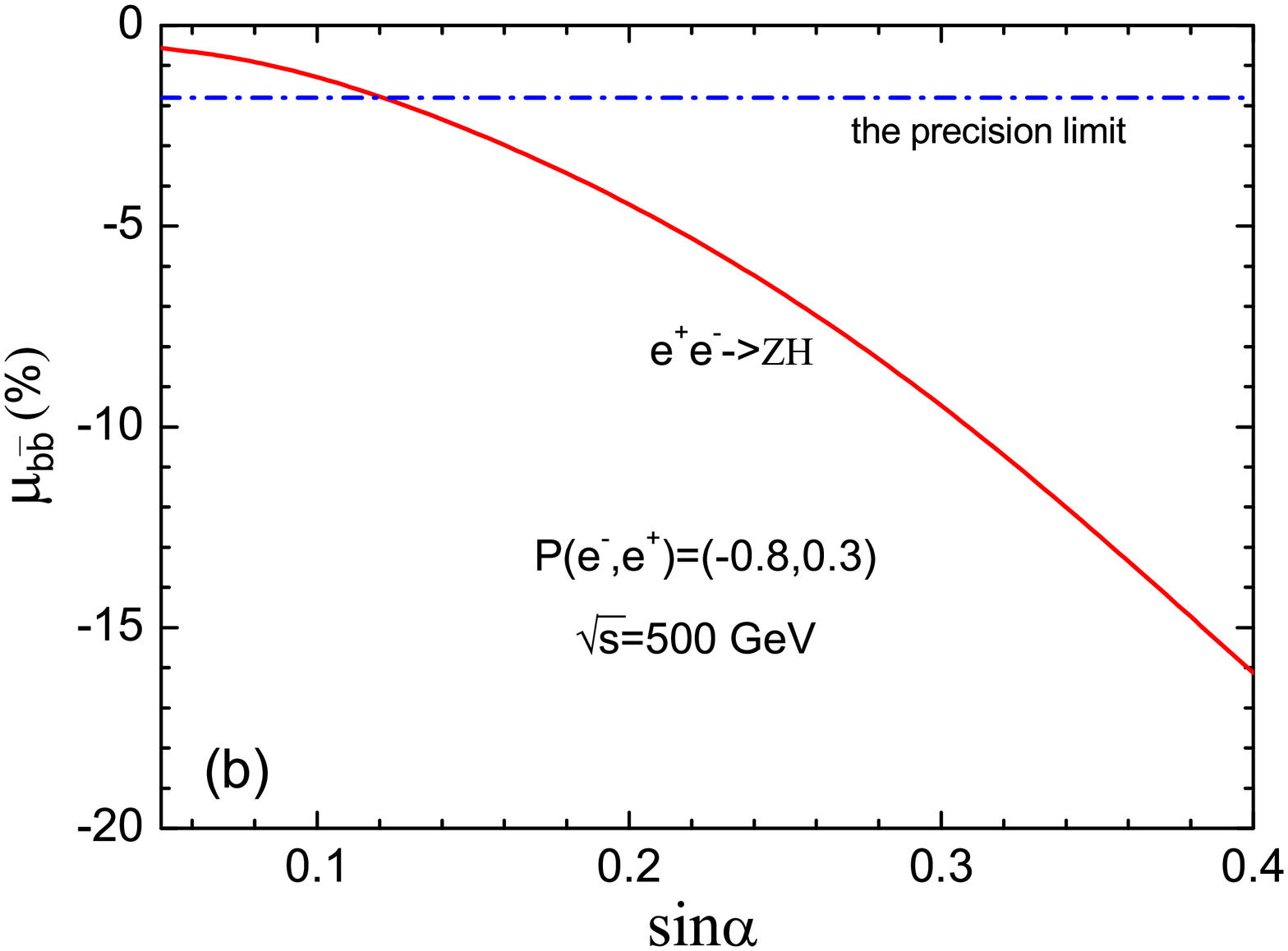}}\vspace{-0.7cm}
\caption{Higgs signal strengths $\mu_{b\bar{b}}$ for the process
$e^{+}e^{-}\rightarrow ZH$ versus ${\rm sin}\alpha$  for
$\sqrt{s}=250$ GeV (a) and $\sqrt{s}=500$ GeV (b) in the $B-L$
model.}\label{fig:ubb-zh}
\end{figure}
\begin{figure}[htbp]
\scalebox{0.3}{\epsfig{file=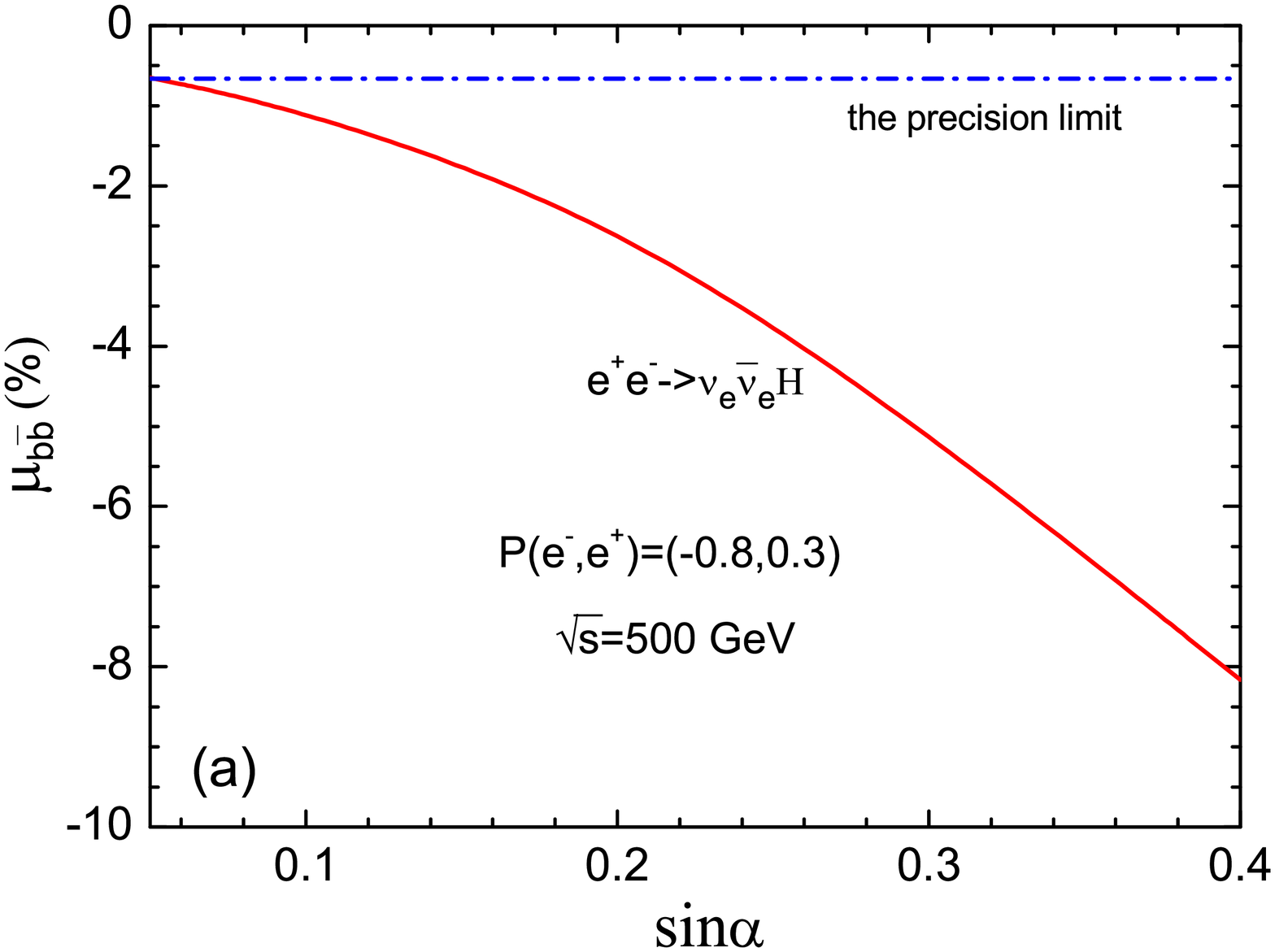} }\hspace{-1.6cm}
\scalebox{0.3}{\epsfig{file=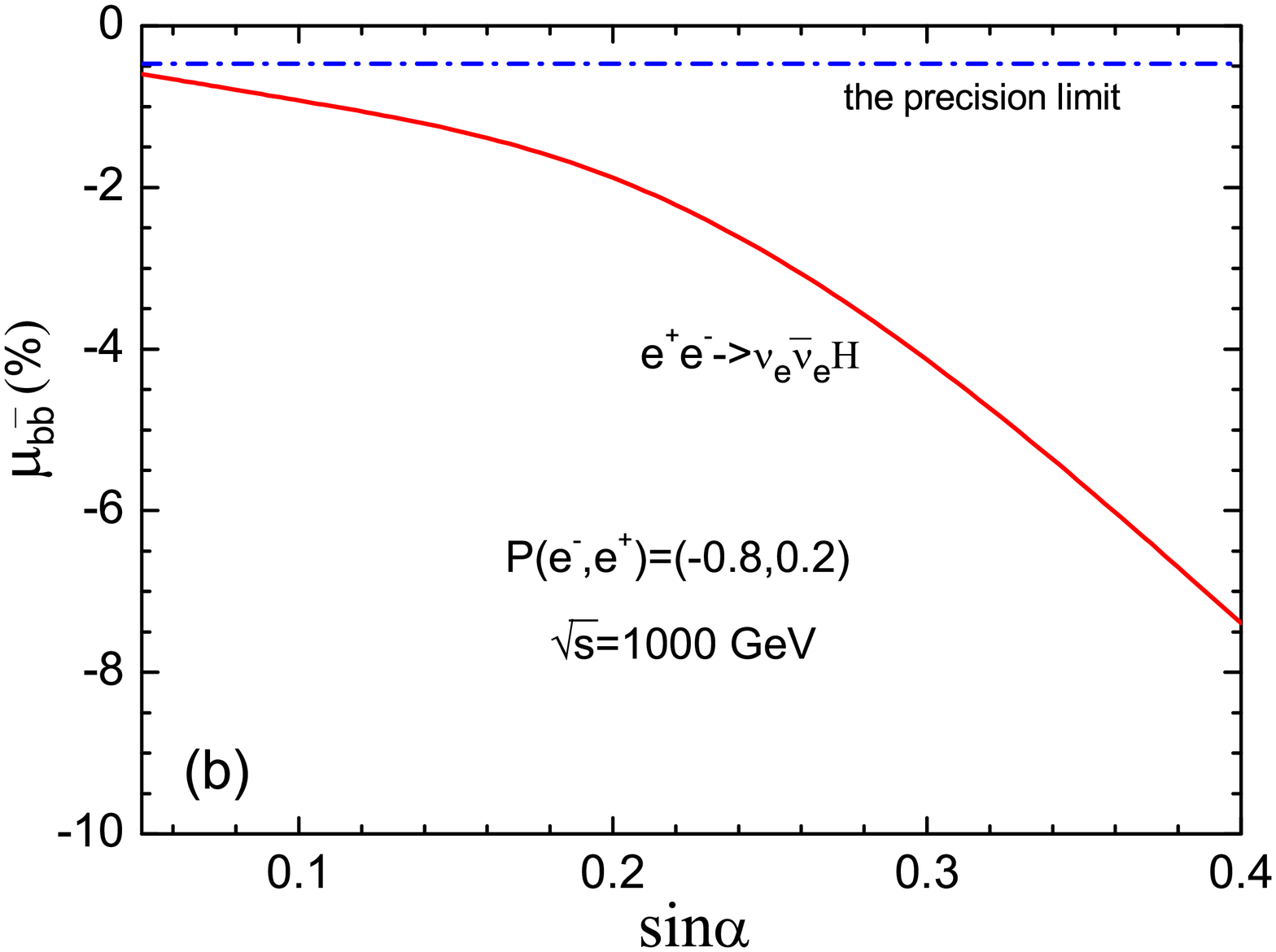} } \vspace{-1.5cm}
\caption{Higgs signal strengths $\mu_{b\bar{b}}$ for the process
$e^{+}e^{-}\rightarrow \nu_{e}\bar{\nu_{e}}H$ versus ${\rm
sin}\alpha$ for $\sqrt{s}=500$ GeV (a) and $\sqrt{s}=1000$ GeV (b)
in the $B-L$ model.}\label{fig:ubb-vvh}
\end{figure}

In Fig.\ref{fig:ubb-zh}, we show the dependence of the Higgs signal
strengths $\mu_{b\bar{b}}$ on the parameter 
sin$\alpha$ for the process $e^{+}e^{-}\rightarrow ZH$ with
polarized beams. From Table \ref{table1}, we can see that the
1.1(1.8)\% accuracy for this mode are expected at $\sqrt{s}$ =
250(500) GeV, and the contributions of the $B-L$ model might be
detected by the measurement of the $b\bar{b}$ signal rate in the
future ILC experiments for sin$\alpha>0.1$.

In Fig.\ref{fig:ubb-vvh}, we show the dependence of the Higgs signal
strengths $\mu_{b\bar{b}}$ on the parameter sin$\alpha$ for the
processes $e^{+}e^{-}\rightarrow \nu_{e}\bar{\nu_{{e}}}H$ with
polarized beams. From Table \ref{table1}, we can see that the
0.66(0.47)\% accuracy for the processes $e^{+}e^{-}\rightarrow
\nu_{e}\bar{\nu_{{e}}}H$ are expected at $\sqrt{s}$ = 500(1000) GeV.
This accuracy is so high that almost any deviation from the SM
prediction can be detected by the measurement of the $b\bar{b}$
signal rate. Conversely, the ILC measurement will give strong bound
on the $B-L$ parameter if this effect can not be detected.

\begin{figure}[htbp]
\scalebox{0.3}{\epsfig{file=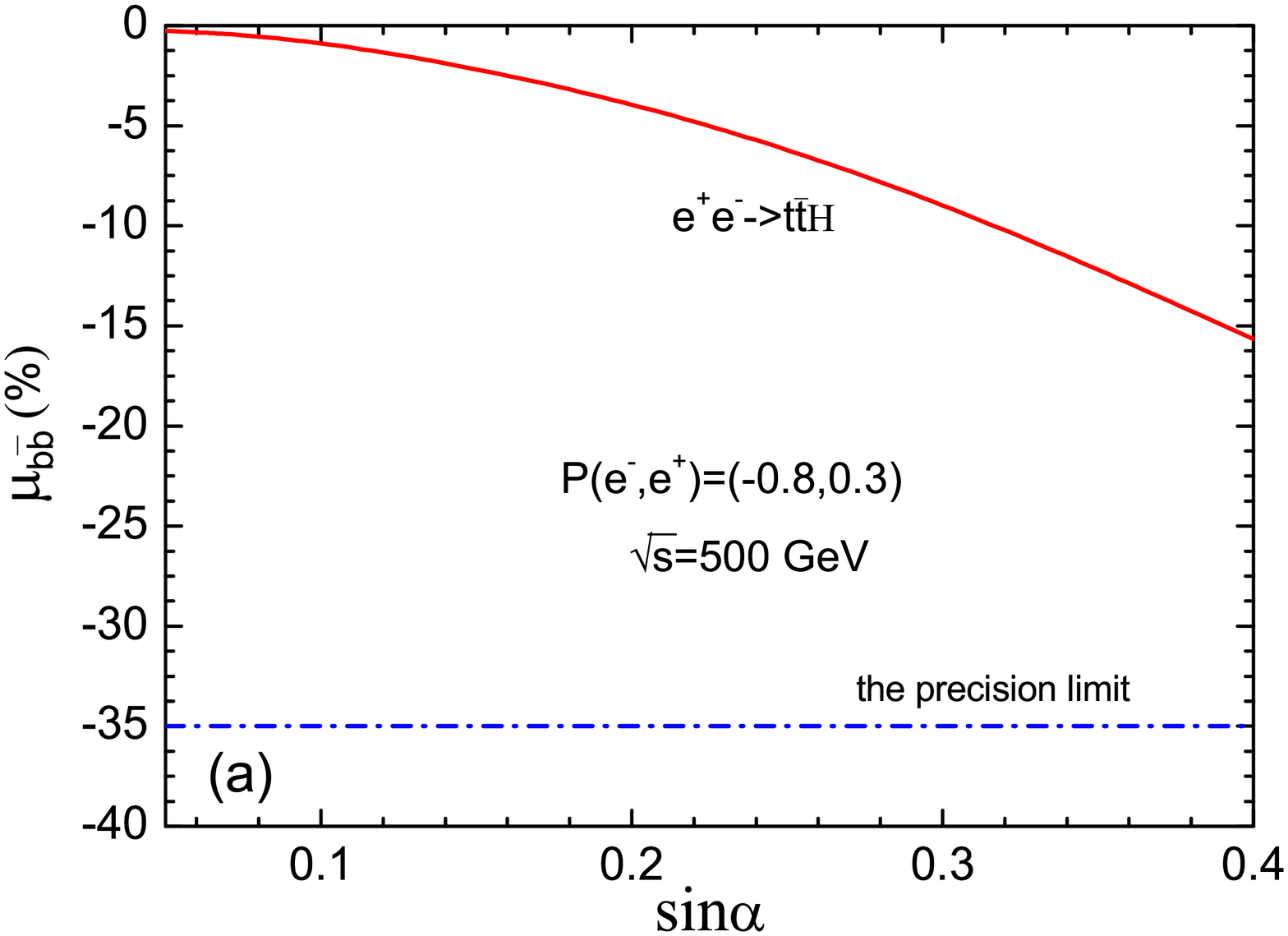}}\hspace{-1.5cm}
\scalebox{0.3}{\epsfig{file=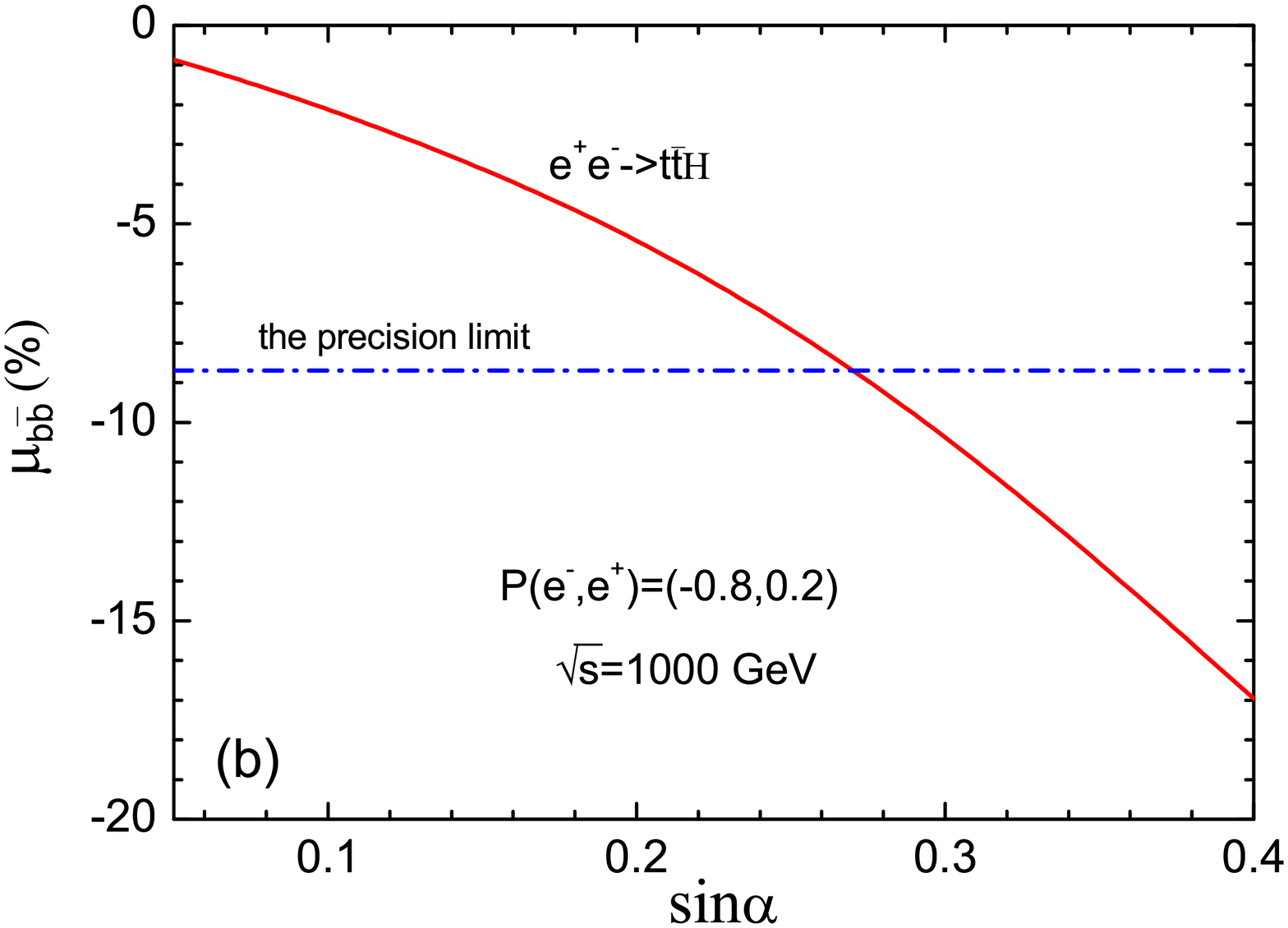}}\vspace{-0.7cm}
\caption{Higgs signal strengths $\mu_{b\bar{b}}$ for the process
$e^{+}e^{-}\rightarrow t\bar{t}H$ versus ${\rm sin}\alpha$ for
$\sqrt{s}=500$ GeV (a) and $\sqrt{s}=1000$ GeV (b) in the $B-L$
model.}\label{fig:ubb-tth}
\end{figure}
\begin{figure}[htbp]
\scalebox{0.3}{\epsfig{file=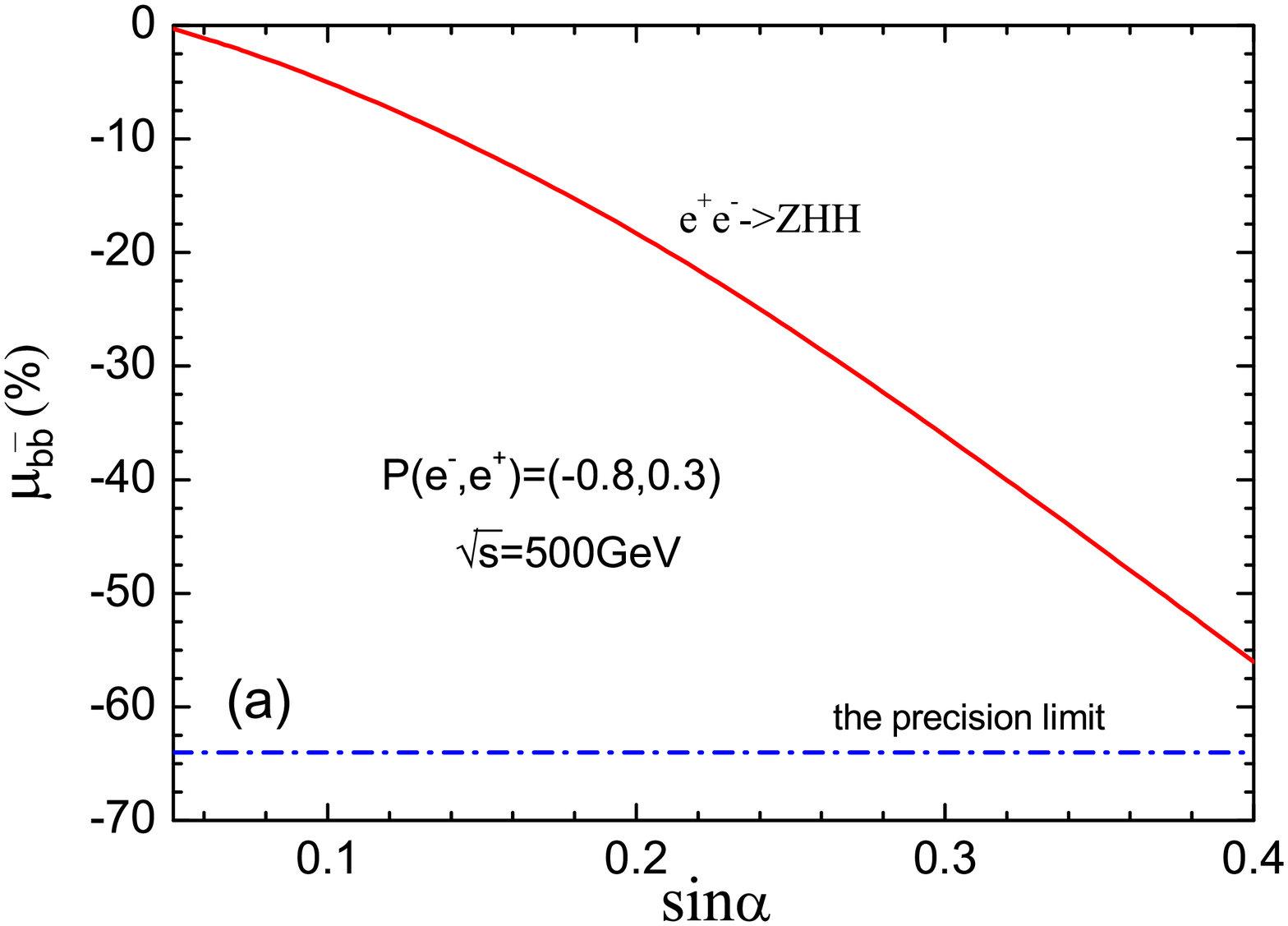}}\hspace{-1.5cm}
\scalebox{0.3}{\epsfig{file=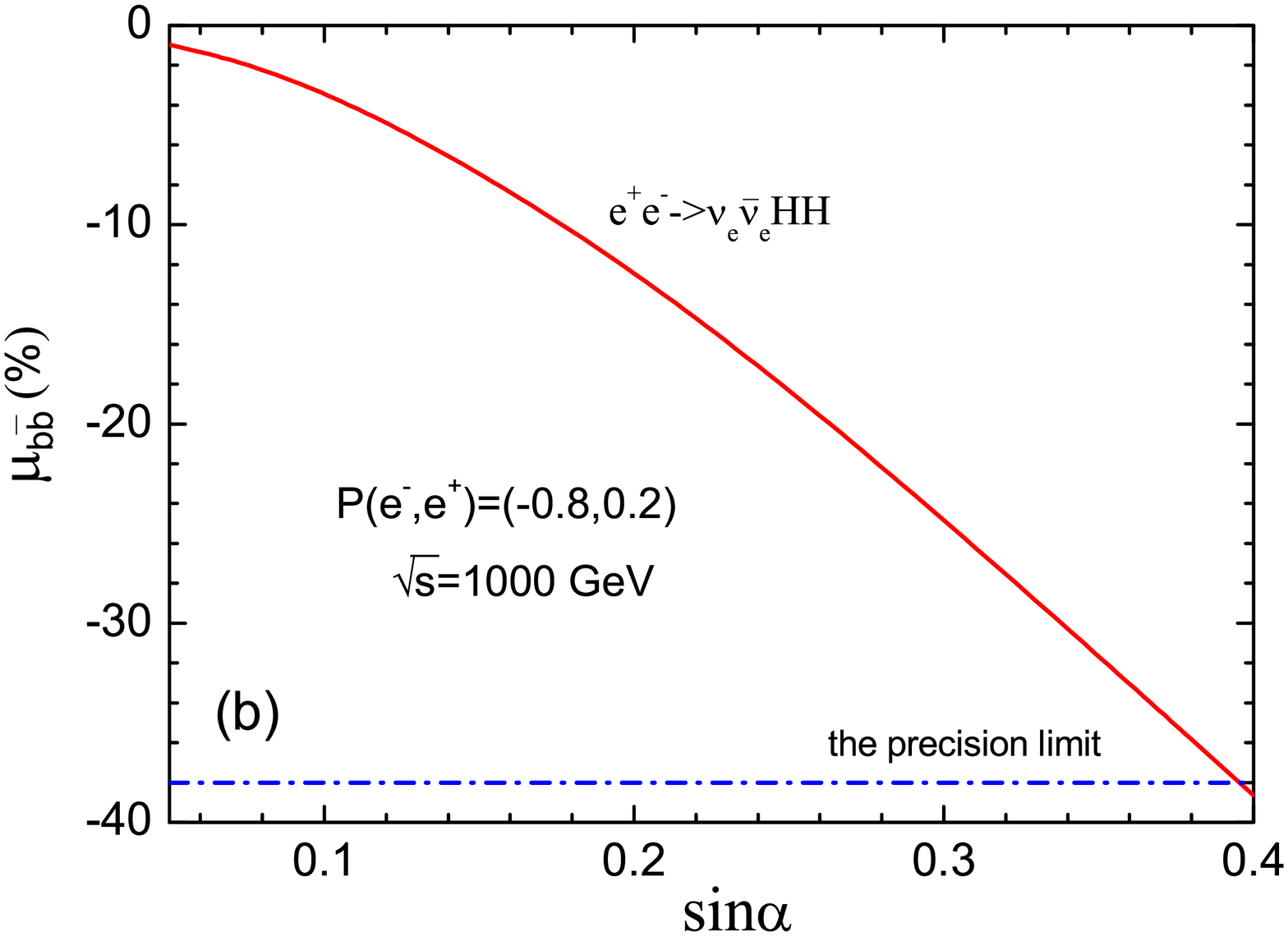}}\vspace{-0.7cm}
\caption{Higgs signal strengths $\mu_{b\bar{b}}$ for the process
$e^{+}e^{-}\rightarrow ZHH$ at $\sqrt{s}=500$ GeV (a) and
$e^{+}e^{-}\rightarrow \nu_{e}\bar{\nu_{e}}HH$ at $\sqrt{s}=1000$
GeV (b) versus ${\rm sin}\alpha$ in the $B-L$
model.}\label{fig:ubb-zhhvvhh}
\end{figure}

In Fig.\ref{fig:ubb-tth}, we show the dependence of the Higgs signal
strengths $\mu_{b\bar{b}}$ on the parameter sin$\alpha$ for the
processes $e^{+}e^{-}\rightarrow t\bar{t}H$ with polarized beams.
From Table \ref{table1}, we can see that the accuracy for top Yukawa
coupling is about 35\% at $\sqrt{s}$ = 500 GeV, which is difficult
to observe the $B-L$ effect on the process $e^{+}e^{-}\rightarrow
t\bar{t}H$ via the $b\bar{b}$ channel. However, this accuracy can be
improved to 8.7\% at $\sqrt{s}$ = 1000 GeV so that the $B-L$ effect
of this process may be detected at the high energy ILC for
sin$\alpha\geq0.3$.

In Fig.\ref{fig:ubb-zhhvvhh}, we show the dependence of the Higgs
signal strengths $\mu_{b\bar{b}}$ on the parameter sin$\alpha$ for
the double Higgs production processes $e^{+}e^{-}\rightarrow ZHH$
and $e^{+}e^{-}\rightarrow \nu_{e}\bar{\nu_{e}}HH$ with polarized
beams, respectively. We can see that the Higgs signal strengths
$\mu_{b\bar{b}}$ of these two processes are both below the expected
precision limits so that these effects will be hard to be observed
at the ILC.
\section{summary}
Under current constraints, we investigated the single and double
Higgs boson production processes $e^{+}e^{-}\rightarrow ZH$,
$e^{+}e^{-}\rightarrow \nu_{e}\bar{\nu_{e}}H$,
$e^{+}e^{-}\rightarrow t\bar{t}H$, $e^{+}e^{-}\rightarrow ZHH$ and
$e^{+}e^{-}\rightarrow \nu_{e}\bar{\nu_{e}}HH$ in the $B-L$  model
at the ILC. We calculated the production cross sections and the
relative corrections with the polarized beams for $\sqrt{s}$=250
GeV, 500 GeV, 1000 GeV. We also studied the signal rates with the
SM-like Higgs boson decaying to $b\bar{b}$, and performed a
simulation by using the projected sensitivities given by the ILC.
For the three single Higgs boson production processes, we found that
the processes $e^{+}e^{-}\rightarrow ZH$ and $e^{+}e^{-}\rightarrow
\nu_{e}\bar{\nu_{e}}H$  might approach the observable threshold of
the ILC in the allowed parameter space. For the two double Higgs
boson production processes, we found that the Higgs signal strengths
$\mu_{b\bar{b}}$ of them are all out of the observed threshold of
the ILC in most regions of parameter space so that the effects will
be difficult to be observed at the ILC.

\section*{Acknowledgement}
We would like to thank Lorenzo Basso and Alexander Belyaev for
providing the CalcHep Model Code and the helpful suggestions. This
work is supported by the Joint Funds of the National Natural Science
Foundation of China under grant No. U1404113, by the National
Natural Science Foundation of China under Grant Nos. 11405047,
11305049, by the China Postdoctoral Science Foundation under Grant
No. 2014M561987.
\newpage
\section*{Appendix}

\begin{table}[ht]
\begin{center}
\caption{ The relevant Feynman rules for single and double Higgs
boson processes in the minimal $B-L$  model at the ILC. }
\label{table2} \vspace{0.2cm}

\begin{tabular}{|l|l|} \hline
vertices & Variational derivative of Lagrangian by fields
\\ \hline
${H}$ \phantom{-} ${Z}_{\mu }$ \phantom{-} ${Z}_{\nu }$ \phantom{-}&$\frac{ c_{\alpha}   e   m_W}{ c_w^2    s_w}g^{\mu \nu} $\\
${H}$ \phantom{-} ${Z}_{\mu }$ \phantom{-} ${Z'}_{\nu}$ \phantom{-}&$\frac{1}{c_wes_w}[s_w^2s_ps_{\alpha}c_pm_W\tilde{g}^2-s_w^4s_pc_{\alpha}c_pm_W\tilde{g}^2$\\${}$& $-s_w c_w c_{\alpha} e m_w\tilde{g}+2s_ws_p^2c_wc_{\alpha}em_w\tilde{g}-s_p  c_{\alpha} c_p e^2 m_w$\\
${}$&$-8s_ws_{\alpha}s_pc_peg'^{2}v'+8s_w^3s_{\alpha}s_pc_peg'^2v']g^{\mu \nu}$ $$\\
${H}$ \phantom{-} ${Z'}_{\mu }$ \phantom{-} ${Z'}_{\nu}$ \phantom{-}&    $-8 s_{\alpha}   g^{\prime}_1{}^2    x  g^{\mu \nu} $\\
${H}$ \phantom{-} $W^+_{\mu }$ \phantom{-} $W^-_{\nu }$ \phantom{-}  & $\frac{ c_{\alpha}   e   m_W}{ s_w}  g^{\mu \nu} $\\
${H'}$ \phantom{-} ${Z}_{\mu }$ \phantom{-} ${Z}_{\nu }$ \phantom{-}& $\frac{ e   m_W   s_{\alpha}}{ c_w^2    s_w}g^{\mu \nu} $\\
$\bar{t}{}$~~~ \phantom{-} $t{}$~~ \phantom{-} ${H}$ \phantom{-} & $-\frac{1}{2}\frac{ c_{\alpha}   e   m_t}{ m_W   s_w}$\\
$\bar{t}{}$~~~ \phantom{-}$t{}$~~ \phantom{-} ${Z}_{\mu}$ \phantom{-} & $-\frac{1}{6}\frac{ e}{ c_w   s_w} \gamma^\mu \big( (3-4 s_w^2)  {(1-\gamma^5)\over 2} -4 s_w^2   {(1+\gamma^5)\over 2} \big)$\\
$\bar{t}{}$~~ \phantom{-} $t{}$~~ \phantom{-} ${Z'}_{\mu}$ \phantom{-}  & $-\frac{1}{3} g^{\prime}_1\gamma^\mu $\\
${\nu l}^i_{a }$ \phantom{-} ${\nu l}^i_{b }$ \phantom{-} ${Z}_{\mu }$ \phantom{-}  & $\frac{1}{2}\frac{ c_{ai}^2    e}{ c_w   s_w}\gamma_{a c}^\mu \gamma_{c b}^5 $\\
${\nu l}^i_{a }$ \phantom{-} ${\nu l}^i_{b }$ \phantom{-}${Z'}_{\mu}$ \phantom{-}  &$- (1-2 s_{ai} ^2)   g^{\prime}_1\gamma_{a c}^\mu \gamma_{c b}^5 $\\
${\nu l}^i_{a }$ \phantom{-} ${\nu h}^1_{b }$ \phantom{-}$~{H}$ \phantom{-}  &  $-\frac{1}{2}\frac{1}{ v'}\big( (1-2 s_{ai} ^2)   c_{\alpha}  \sqrt{2}   v'   y^{\nu}_i  \delta_{a b} +2 s_{\alpha}   s_{ai}   c_{ai}   m_{\nu_i}  \delta_{a b} \big)$\\
${\nu l}^i_{a }$ \phantom{-} ${\nu h}^1_{b }$ \phantom{-}${H'}$ \phantom{-}  & $-\frac{1}{2}\frac{1}{ v'}\big( (1-2 s_{ai} ^2)   s_{\alpha}  \sqrt{2}   v'   y^{\nu}_i  \delta_{a b} -2 s_{ai}   c_{\alpha}   c_{ai}   m_{\nu_i}  \delta_{a b} \big)$\\
${H}$~~\phantom{-} ${H}$~\phantom{-} ${H}$ \phantom{-}  & $-3\frac{1}{ e}\big(4 c_{\alpha}^3  s_w m_W \lambda _1-2 s_{\alpha}^3  e \lambda _2 v'- c_{\alpha}^2  s_{\alpha} e \lambda _3 v'$ \\
  & $+2 s_w s_{\alpha}^2  c_{\alpha} m_W \lambda _3\big)$\\
${H}$ \phantom{-}~~${H}$~~\phantom{-} ${H'}$ \phantom{-}  &
    $-\frac{1}{ e}\big(12 c_{\alpha}^2  s_w s_{\alpha} m_W \lambda _1+6 s_{\alpha}^2  c_{\alpha} e \lambda _2 v'+ (1-3 s_{\alpha}^2) c_{\alpha} e \lambda _3 v'$ \\
  & $-2 (2-3 s_{\alpha}^2) s_w s_{\alpha} m_W \lambda _3\big)$\\
${H}$ \phantom{-} ${H}$ \phantom{-} $W^+_{\mu }$ \phantom{-}
$W^-_{\nu }$ \phantom{-}  &$\frac{1}{2}\frac{ c_{\alpha}^2    e^2 }{ s_w^2 }  g^{\mu \nu} $\\
${H}$ \phantom{-} ${H}$ \phantom{-} ${Z}_{\mu }$ \phantom{-}
${Z}_{\nu }$ \phantom{-}  &$\frac{1}{2}\frac{ c_{\alpha}^2    e^2 }{ c_w^2    s_w^2 }g^{\mu \nu} $\\
 \hline

\end{tabular}
\end{center}
\end{table}

 Here, $e$ is the electric charge,
 $s_w$($c_w$) $\Rightarrow$ $\sin{\theta_W}$($\cos{\theta_W}$),
$s_\alpha$($c_\alpha$) $\Rightarrow$ $\sin{\alpha}$($\cos{\alpha}$),
$s_{\alpha i}$($c_{\alpha i}$) is the sinus(cosinus) of the
  ``see-saw'' mixing of the $i^{th}$ neutrino generation,
  $s_p=\frac{1}{2}{\rm sin(arcsin}(s_n/\sqrt{s_n^2+c_n^2}))$, $c_p=\sqrt{1-s_p^2}$,
  $s_n=2\tilde{g}\sqrt{(\frac{e}{s_w})^2+(\frac{e}{c_w})^2}$,
  $c_n=\tilde{g}^2+16g'^2(\frac{v'}{v})^2-(\frac{e}{s_w})^2-(\frac{e}{c_w})^2$.


\newpage
\begin{thebibliography}\\
\bibitem{ATLAS-H}
    ATLAS Collaboration (G. Aad {\it et al}.), \PLB716, 1-29 (2012).

\bibitem{CMS-H}
    CMS Collaboration (S. Chatrchyan {\it et al}.), \PLB716, 30-61 (2012).

\bibitem{LHC-1}
   T. Han, Z. Liu and J. Sayre, \PRD89, 113006 (2014).

\bibitem{LHC-2}
   P. Bechtle, S. Heinemeyer, O. Stal, T. Stefaniak and G. Weiglein, \JHEP1411, 039 (2014).

 \bibitem{LHC-3}
   C. Englert, A. Freitas, M. Muhlleitner {\it et al}., \JPG41, 113001 (2014).

 \bibitem{LHC-4}
   M. E. Peskin, arXiv:1312.4974 [hep-ph].

\bibitem{LHC-5}
   S. Dawson, A. Gritsan, H. Logan {\it et al}., arXiv:1310.8361 [hep-ex].

\bibitem{ILC-1}
   T. Behnke, J. E. Brau, B. Foster {\it et al}., arXiv:1306.6327 [acc-ph].
\bibitem{ILC-2}
   H. Baer, T. Barklow, K. Fujii {\it et al}., arXiv:1306.6352 [hep-ph].

\bibitem{ILC-3}
 D. M. Asner, T. Barklow, C. Calancha {\it et al}., arXiv:1310.0763 [hep-ph].

\bibitem{SM-zh-vvh-eeh-1}
   B. A. Kniehl, \IJMPA17, 1457-1476 (2002).

\bibitem{SM-zh-vvh-eeh-2}
   F. Jegerlehner, O. Tarasov, \NPPS116, 83-87 (2003).

\bibitem{SM-zh-vvh-eeh-3}
   G. Belanger {\it et al}., \PLB559, 252-262 (2003).

\bibitem{SM-zh-vvh-eeh-4}
   F. Boudjema {\it et al}., \PLB600, 65-76 (2004).

\bibitem{SM-zh-vvh-eeh-5}
   A. Denner, S. Dittmaier, M. Roth and M. M. Weber, \PLB560,  196-203 (2003).

\bibitem{SM-zh-vvh-eeh-6}
   A. Denner, S. Dittmaier, M. Roth and M. M. Weber, \NPB660, 289-321 (2003).

\bibitem{SM-zh-vvh-eeh-7}
   P. S. Bhupal Dev {\it et al}., \PRL100, 051801 (2008).

\bibitem{np-1}
   H. Eberl, W. Majerotto and V. C. Spanos, \NPB657, 378-396 (2003).

\bibitem{np-2}
   T. Hahn, S. Heinemeyer and G. Weiglein, \NPB652,  229-258 (2003).

 \bibitem{np-3}
   J. J. Cao, C. C. Han, J. Ren {\it et al}., arXiv:1410.1018 [hep-ph].

\bibitem{np-4}
   A. GutiL\'{e}rrez-RodrL\'{i}guez and M. A. Hern\'{a}ndez-Ruiz, {\it Adv. High Energy Phys.} {\bf 2015}, 593898 (2015).

\bibitem{np-5}
   S. Banerjee {\it et al}., \PRD92,  075002 (2015).

\bibitem{np-6}
   C. -X. Yue, S. Z. Wang and D. Q. Yu, \PRD68,  115004 (2003).

\bibitem{np-7}
   C. -X. Yue, W. Wang, Z. J. Zong and F. Zhang, \EPJC42,  331 (2005).

\bibitem{np-8}
   X. L. Wang, Y. -B. Liu, J. H. Chen and H. Yang, \EPJC49, 593-597 (2007).

\bibitem{np-9}
   S. L. Hu, N. Liu, J. Ren and L. Wu, \JPG41, 125004 (2014).

\bibitem{np-10}
   N. Liu, J. Ren, L. Wu , P. W. Wu and J. M. Yang, \JHEP1404,  189 (2014).
\bibitem{np-11}
   B. F. Yang, J. Z. Han, S. H. Zhou and N. Liu, \JPG41, 075009 (2014).

\bibitem{np-12}
    S. Antusch, E. Cazzato and O. Fischer,  \JHEP1604, 189 (2016).

\bibitem{np-13}
   L. Wang, W. Y. Wang, J. M, Yang and H. J. Zhang, \PRD75, 074006 (2007).

\bibitem{np-14}
   J. F. Shen, J. Cao and L. B. Yan, \EPL91, 51001 (2010).

\bibitem{np-15}
   N. Liu, S. L. Hu, B. F. Yang and J. Z. Han, \JHEP1501, 008 (2015).

\bibitem{np-16}
   B. F. Yang,  Z. Y. Liu, N. Liu and J. Z. Han, \EPJC74, 3203 (2014).

\bibitem{np-17}
   B. F. Yang, J. Z. Han and N. Liu, \JHEP1504,  148 (2015).

\bibitem{np-18}
    L. Wu, \JHEP1502, 061 (2015).

\bibitem{np-19}
   Y.-B. Liu and Z.-J. Xiao, \JPG42, 065005 (2015).

\bibitem{np-20}
   J. Z. Han,  S. F. Li, B. F. Yan and N. Liu, \NPB896,  200-211 (2015).

\bibitem{B-L-1}
   S. Khalil, \JPG35, 055001 (2008).

\bibitem{B-L-2}
   L. Basso, arXiv:1106.4462.

\bibitem{B-L-LHC-ILC-1}
   W. Emam and S. Khalil, \EPJC52, 625-633 (2007).

\bibitem{B-L-LHC-ILC-2}
   L. Basso, A. Belyaev, S. Moretti and C. H. Shepherd-Themistocleous, \PRD80, 055030 (2009).

\bibitem{B-L-LHC-ILC-3}
   L. Basso, A. Belyaev, S. Moretti and G. M. Pruna, \JHEP0910, 006 (2009).

\bibitem{B-L-LHC-ILC-4}
   P.~Fileviez~Perez, T.~Han, and T.~Li, \PRD80, 073015 (2009).


\bibitem{B-L-LHC-ILC-5}
   L. Basso, S. Moretti, and G. M. Pruna, \EPJC71,  1724 (2011).

 \bibitem{B-L-LHC-ILC-6}
    L. Basso, S. Moretti and G. M. Pruna, \PRD83, 055014 (2011).

 \bibitem{B-L-LHC-ILC-7}
   G. M. Pruna, arXiv:1106.4691 [hep-ph].

\bibitem{B-L-LHC-ILC-8}
   C. Englert, T. Plehn, D. Zerwas and P. M. Zerwas, \PLB703, 298-305 (2011).

\bibitem{B-L-LHC-ILC-9}
    V. V. Khoze and G. Ro, \JHEP1310, 075 (2013).

\bibitem{B-L-LHC-ILC-10}
   J. Hern\'{a}ndez~L\'{o}pez and J. Orduz-Ducuara, {\em J. Phys. Conf. Ser.} {\bf 468},  012012 (2013).

\bibitem{B-L-extension-1}
   R.~Marshak and R.~N. Mohapatra, \PLB91,  222-224 (1980).

\bibitem{B-L-extension-2}
  R. N. Mohapatra and R.~Marshak, \PRL44, 1316-1319 (1980).

\bibitem{B-L-extension-3}
   C. Wetterich, \NPB187, 343-375 (1981).

\bibitem{B-L-extension-4}
   A. Masiero, J. Nieves and T. Yanagida, \PLB116, 11-15 (1982).

\bibitem{B-L-extension-5}
   R. N. Mohapatra and G. Senjanovic, \PRD27, 254 (1983).

\bibitem{LEP-constrain}
   G. Cacciapaglia, C. Csaki, G. Marandella and A. Strumia, \PRD74, 033011 (2006).


\bibitem{PDG-2014}
   Particle Data Group collaboration (K. A. Olive {\it et al}.), \CPC38, 090001 (2014).

\bibitem{B-L-consition-1}
   S. Banerjee, M. Mitra and M. Spannowsky, \PRD92, 055013 (2015).

\bibitem{calchep}
   A. Belyaev, N. D. Christensen and A. Pukhov, {\it Comput. Phys. Commun.} {\bf 184}, 1729 (2013).

\bibitem{polarization-1}
   G. Moortgat-Pick, T. Abe, G. Alexander {\it et al}., \PR460, 131 (2008).

\bibitem{polarization-2}
   J. J. Cao, Z. X. Heng, L. Wu and J. M. Yang, \PRD81, 014016 (2010).

\bibitem{sm1}
   A. Denner, J. Kublbeck, R. Mertig and M. Bohm, \ZPC56, 261 (1992).

\bibitem{sm2}
   C. Englert and M. McCullough, \JHEP1307, 168 (2013).

\bibitem{HL-LHC-1}
   M. E. Peskin, arXiv:1312.4974 [hep-ph].

\bibitem{HL-LHC-2}
   F. Goertz, A. Papaefstathiou, L. L. Yang and J. Zurita, \JHEP1306, 016 (2013).

\bibitem{HL-LHC-3}
   R. S. Gupta, H. Rzehak and J. D. Wells, \PRD88,  055024 (2013).
\bibitem{HL-LHC-4}
   A. J. Barr, M. J. Dolan, C. Englert and M. Spannowsky, \PLB728,  308-313 (2014).

\bibitem{HL-LHC-5}
   V. Barger, L. L. Everett, C. B. Jackson and G. Shaughnessy, \PLB728,  433 (2014).
\bibitem{HL-LHC-6}
   D. E. F. de Lima, A. Papaefstathiou and M. Spannowsky, \JHEP1408, 030 (2014).
\end{thebibliography}
\end{document}